\documentclass[nofootinbib,aps, preprint, floatfix]{revtex4}

\input{epsf}
\usepackage{graphicx}

\begin{document}
\vspace{3.0cm}
\preprint{\vbox {\vspace{2cm}
\hbox{WSU--HEP--0905}}}
\vspace*{4cm}

\title{$\Upsilon$ Decays into Light Scalar Dark Matter}

\author{Gagik K. Yeghiyan\vspace{5pt}}
\email{ye_gagik@wayne.edu}
\affiliation{Department of Physics and Astronomy\\
        Wayne State University, Detroit, MI 48201\vspace{15pt}}

\begin{abstract}
\vspace{5pt}
We examine decays of a spin-1 bottomonium into a pair of
light scalar Dark Matter (DM) particles, assuming that Dark
Matter is produced due to exchange of heavy
degrees of freedom.
We perform a model-independent
analysis
and derive formulae for the branching ratios of these decays. We
confront our calculation results with the experimental data.
We show that the considered branching ratios are within the
reach of the present BaBaR experimental sensitivity. Thus, Dark Matter
production in $\Upsilon$ decays leads to
constraints on parameters of various models containing a
light spin-0 DM particle. We illustrate this for the models with a
"WIMPless miracle", in particular for a
Gauge Mediated SUSY breaking
scenario, with a spin-0 DM particle in the hidden sector. Another example
considered is the type~II 2HDM with a scalar DM particle.
\end{abstract}
\maketitle
\renewcommand{\thesection}{\arabic{section}}

\section{Introduction}\label{s1}
\renewcommand{\theequation}{\ref{s1}.\arabic{equation}}
\setcounter{equation}{0}

The origin, presence and nature of Dark Matter (DM) in our Universe
remains one of the biggest mysteries of the particle physics and astrophysics \cite{51}.
Understanding the nature of Dark Matter, which accounts for majority of mass
in the Universe,
represents a crucial step in connecting astronomical observations with predictions
of various elementary particle theories.

Many theories, the extensions of the Standard
Model (SM), predict one or more stable, electrically-neutral particle(s) in their spectrum,
which can be possible Dark Matter candidate(s).
Different models provide different assignments for DM particles' spin and various
windows for their masses and couplings. To test this great variety of hypotheses, several
techniques for DM direct or indirect search are currently developed.

Recent experimental measurements of Dark Matter relic
abundance, $\Omega_{DM} h^2 \sim 0.11$ by WMAP collaboration~\cite{46},
can be used to place constraints on the masses and interaction strengths of DM
particles. Indeed, the relation
\begin{equation}\label{i1}
\Omega_{DM} h^2 \sim \langle \sigma_{ann} v_{rel}\rangle^{-1} \propto \frac{M^2}{g^4},
\end{equation}
with $M$ and $g$ being the mass and the interaction strength associated with DM
annihilation, implies that, for a weakly-interacting particle, the mass scale should be
set around the electroweak scale.
This, coupled with an observation that very light DM
particles might overclose the Universe  (known as a Lee-Weinberg limit~\cite{47}) ,
seems to exclude the possibility of light weakly-interacting massive particle (WIMP)
solution for DM, setting $M_{DM} > 2-6$~GeV.

A detailed look at this argument reveals that those constraints can be easily
avoided in concrete models, so even MeV-scale particles can be good candidates for
DM. For instance, low-energy resonances - such a light CP-odd Higgs in the MSSM
extensions with a Higgs singlet, or a light
extra gauge U-boson - could enhance the DM annihilation
cross-section, without the need for a large coupling constant \cite{48}, \cite{24},
\cite{34}-\cite{50}.
Even if no light resonances exist,
the usual suppression of DM annihilation cross-section, $M^4_{DM}/M^4$, used in setting the
Lee-Weinberg limit, does not hold, if Dark Matter is a non-fermionic (e.g. spin-0)
state \cite{34,2,16,15,68,69,70}.
Furthermore,
DM production could be non-thermal ~\cite{52,67}, in which case the constraint provided by
Eq.~(\ref{i1}) does not apply.  Thus, models with light
Dark Matter ($M_{DM} \sim$ few GeV or less) still deserve a detailed and thorough study.

In this paper we consider the possibility of using of
$\Upsilon$ meson decays with missing energy,
to test the models with a light spin-0 DM particle. Studies of heavy meson (and in
particular $\Upsilon$ meson) decays with missing energy may be especially
valuable in light of the fact that DM direct
search experiments, such as DAMA \cite{54,55,56},
CDMS \cite{57} and XENON \cite{58,59}, which rely on the measurement of
kinematic recoil of nuclei in DM interactions, lose (for cold DM particles)
sensitivity with decreasing mass of the WIMP, as the recoil energy becomes small.
Of course, light Dark Matter may also be produced at high energy colliders, however the
production rate is naturally going to be insensitive to precise value of the WIMP mass,
if the one is much less than the beam center-of-mass energy \cite{53}.
Indirect
experiments, such as HESS \cite{60}, are specifically tuned to see large energy secondaries, only
possible for weak-scale WIMPs. The backgrounds for positron and antiproton searches
by HEAT \cite{63}
and/or PAMELA \cite{61,62} experiments could be prohibitively large at small energies. Thus,
$\Upsilon$ (and/or other heavy meson) decays with missing energy may serve as alternative DM
search channels, capable to provide us with an information on the WIMP mass range, hardly testable
by the experiments discussed above.

So far, $\Upsilon$ meson decays into Dark Matter have been considered within the models,
where DM particles interaction with an ordinary matter is
mediated by some light degree of freedom \cite{48,37,36}. Apart from desire of having DM
annihilation enhancement (due to a light intermediate resonance) and thus having no
tension with the DM relic abundance condition, it is also known that
$\Upsilon$ meson SM decay is predominantly due to strong interactions. Thus,
the
WIMP production branching ratio, in general, is greatly suppressed compared to
relevant weak B decays, and in particular
to $B \to K + invisible$ transition \cite{16,15}. In light of this,
it might seem natural to concentrate only on the models within
which Dark Matter production in $\Upsilon$ decays is enhanced due to
exchange of a light particle propagator.

Yet, the aim of the present paper
is to study $\Upsilon(1S)$ decay into a pair of spin-0 DM particles,
$\Upsilon(1S) \to \Phi \Phi^*$, and  $\Upsilon(3S)$ decay into a pair of
spin-0 DM particles and a photon, $\Upsilon(3S) \to \Phi \Phi^* \gamma$,
within the models
where light Dark Matter interaction with an ordinary matter is due to exchange of
{\it heavy} particles (with masses exceeding the bottomonium mass).
As it is mentioned above, these models may
be free of tension related to satisfying
the DM relic abundance constraint as well.
Examples of such models will be discussed below.
Also, new experimental data
on $\Upsilon$ decays into invisible states have been reported by the BaBaR
collaboration \cite{66,21}. According to these data,
\begin{equation}
B(\Upsilon(1S) \to invisible) < 3 \times 10^{-4} \label{i2}
\end{equation}
and
\begin{equation}
B(\Upsilon(3S) \to \gamma + invisible) < (0.7 - 31) \times 10^{-6} \label{i3}
\end{equation}
where the interval in the r.h.s. of Eq.~(\ref{i3}) is related to the choice of the final state missing mass.
These bounds are significantly stronger than those on invisible
$\Upsilon(1S)$ decays (with or without  a photon emission), reported previously
by Belle and CLEO \cite{64,65} and quoted by Particle Data Group \cite{1}. We show here
that BaBaR
experimental data on $\Upsilon$ meson invisible decays (without or with a photon emission)
may constrain
the parameter space of light scalar Dark Matter models, even if
there is no Dark Matter production
enhancement due to light intermediate states.

We also illustrate (in Sections~\ref{s5}~and~\ref{s4}) that the study of
Dark Matter production in $\Upsilon$ decays allows us to test regions of
parameter space of
light spin-0 DM models that are inaccessible for B meson
decays with missing energy. It is also worth
mentioning that $\Upsilon$ decays are sensitive to
a wider range of WIMP mass than B decays. Thus, the study of
WIMP production in $\Upsilon$
decays is complementary to that
for B meson decays.

Within the  models where scalar DM consists of
particles that are their own antiparticles,  only
$\Upsilon \to \Phi \Phi \gamma$ transition is relevant. The transition rate of
$\Upsilon \to \Phi \Phi^*$  vanishes, if $\Phi = \Phi^*$.
Indeed, by angular momentum conservation, the produced DM particles pair in
$\Upsilon \to \Phi \Phi$ decay must be in a P-wave state, which
is impossible because of the
Bose-Einstein symmetry of identical spin-0 particles.

The scenarios with light complex scalar Dark Matter, albeit implying some continuous
symmetry related to the internal charge of the complex (electrically) neutral scalar field,
may be
realized in many extensions of the Standard Model. Some of the models allow the
scenarios both with a real and with a complex light scalar DM field
\cite{34,68,69,70,35}. Within these models, study of the decay
$\Upsilon(1S) \to \Phi \Phi^*$
represents an excellent opportunity to test the DM field nature
either at present or in the future (if higher experimental resolution
is needed).

As mentioned above, we restrict ourselves to the class of models
where light spin-0
DM production is mediated by heavy degrees of freedom.
At the energy scales, associated with $\Upsilon$ decays,
heavy intermediate degrees of freedom may be
integrated out, thus leading to a low-energy effective theory of four-particle
interactions.
Our strategy would be deriving first model-independent formulae for the
$\Upsilon(1S) \to
\Phi \Phi^*$ and $\Upsilon(3S) \to \Phi \Phi^* \gamma$ branching ratios
within the low-energy effective theory.
Then, we confront our predictions with the experimental data, deriving
model-independent bounds in terms of the Wilson operator expansion
coefficients as the parameters that carry the information on an
underlying New Physics (NP) model. Finally,
within a given model, using the matching conditions for the Wilson coefficients,
we translate these bounds into those on the relevant parameters of the considered
model.

The paper is organized as follows.
In Section~\ref{s2} the general formalism is developed and
model-independent formulae for the $\Upsilon(1S) \to
\Phi \Phi^*$ and $\Upsilon(3S) \to \Phi \Phi^* \gamma$
branching ratios are derived. Both the
case of complex and the case of self-conjugate DM field are considered.
In Section~\ref{s7}
the neutrino background effect is discussed.
We show that it has negligible impact on
our analysis. In Section~\ref{s8} we confront our calculation results
with the experimental data and derive model-independent bounds on the
Wilson coefficients. In the next sections we transform
these bounds into constraints on the parameters of particular models.
In Section~\ref{s5}
models with a complex spin-0 DM field are considered.
We choose the class of mirror fermion
models as an example.
These models include, in particular,
Gauge Mediated SUSY Breaking scenarios with the
DM particle in the hidden sector and the  mirror fermions
being connectors between the
hidden and the MSSM sectors \cite{68,69,70}.
An example of the self-conjugate DM
scenario,
two-Higgs doublet model with a scalar DM particle,
is considered in Section~\ref{s4}.
The concluding remarks are given in Section~\ref{s3}.

\section{General Formalism: Model-Independent Formulae
for the Decays Branching Ratios}\label{s2}

\renewcommand{\theequation}{\ref{s2}.\arabic{equation}}
\setcounter{equation}{0}
We treat $\Upsilon$ states - neglecting the sea quark and gluon
distributions - as
bound states of $b\bar{b}$ valence quark-antiquark pair that
annihilates - with or without emission of a photon - into a pair of
Dark Matter particles. To this approximation,
the relevant low-energy effective Hamiltonian may be written as
\begin{equation}
H_{eff} = \frac{2}{\Lambda_H^2} \sum_i{C_i \ O_i} \label{g1}
\end{equation}
where $\Lambda_H$ is the heavy mass  and
\newpage
\begin{eqnarray}
\nonumber
&& O_1 =  m_b \left(\bar{b} \ b \right) \left( \Phi^* \Phi \right),
\hspace{0.9cm} O_2 = i m_b \left(\bar{b} \gamma_5 b\right) \left(
\Phi^* \Phi \right), \\
&& O_3 = \left(\bar{b} \gamma^\mu  b \right)
\left(\Phi^* i \partial^{^{^{\hspace{-0.22cm} \leftrightarrow}}}_\mu
\Phi \right), \hspace{0.5cm} O_4 = \left(\bar{b} \gamma^\mu
\gamma_5 b \right) \left(\Phi^* i
\partial^{^{^{\hspace{-0.22cm} \leftrightarrow}}}_\mu \Phi \right)
\label{g2}
\end{eqnarray}
with $\partial^{^{^{\hspace{-0.22cm} \leftrightarrow}}} = 1/2
(\overrightarrow{\partial} - \overleftarrow{\partial})$.
It is worth noting that with the notations used in (\ref{g1})
and (\ref{g2}), all the operators $O_i$, i=1,..4, are Hermitean, thus all the
Wilson coefficients $C_i$ must be real. Another point to be made is that
the $C_i$ are low-energy renormalization
scale-independent. This stems from the renormalization scale invariance of the
hadronic parts of
operators $O_i$, which - combined with the fact that DM particles
do not interact strongly or electromagnetically - leads to low-energy scale invariance of
$O_i$ and (from the scale invariance of the effective Hamiltonian) to that
of $C_i$.
If DM consists of
particles that are their own antiparticles, then only first two operators in Eq.~(\ref{g2})
would contribute.

\begin{figure}[t]
\vspace{-0.5cm}
\includegraphics{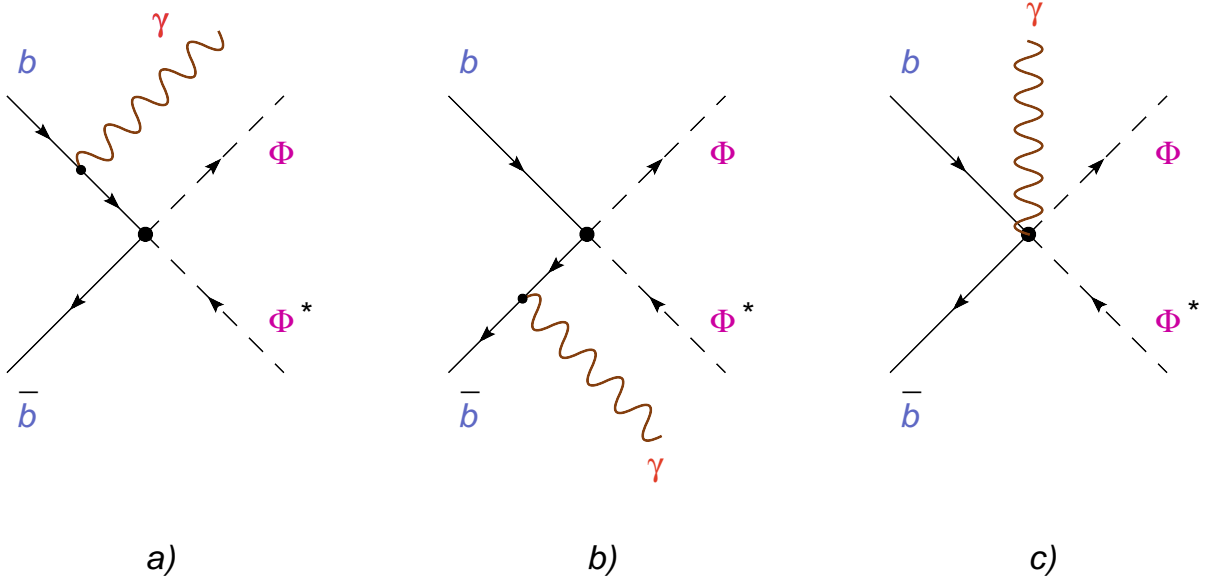}
\caption{Diagrams for $\Upsilon(3S) \to \Phi \Phi^* \gamma$ transition:
a), b) transition is generated by a bi-local interaction c) transition is
generated by an effective local interaction.}
\label{f1}
\end{figure}

To the considered approximation, the operator basis, given by Eq.~(\ref{g2}), is
the most general one for the $\Upsilon(1S) \to \Phi \Phi^*$ transition.
With the use of the same basis, the other decay,
$\Upsilon(3S) \to \Phi \Phi^* \gamma$, occurs by means of bi-local
interactions depicted in Fig.~\ref{f1}~(a),~(b). In principle, it is also
possible - if the intermediate heavy states are (electrically) charged -
that both the photon and the DM particles originate from the same
effective vertex (Fig.~\ref{f1}~(c)). In order to take into account this diagram,
higher dimension operators, involving electromagnetic field tensor, should
also be included. However, such higher dimension
operators would enter the expression for
$H_{eff}$ with higher powers of $1/\Lambda_H$ factor.  Extra suppression of diagram
in Fig.~\ref{f1}(c) in powers of the heavy mass inverse
may be seen from the fact that one must introduce
an extra heavy propagator or, equivalently, an extra
power of $1/\Lambda_H$ or $1/\Lambda_H^2$,
as the photon is emitted
by a heavy intermediate degree of freedom.

Thus, one may neglect contribution of the diagram in Fig.~\ref{f1}(c), as
compared to that of the other two diagrams. This justifies the use of  the operators basis,
given by Eq.~(\ref{g2}), for $\Upsilon(3S) \to \Phi \Phi^* \gamma$ decay as well.

In the rest frame of $\Upsilon$ meson, matrix elements of the local
hadronic currents may be parameterized at the origin, $x = 0$, as
\begin{eqnarray}
\nonumber
&& \hspace{-1.5cm}
\langle 0 | \ \bar{b}(0)  b(0) \ | \Upsilon \rangle =
\langle 0 | \ \bar{b}(0) \gamma_5 b(0) \ | \Upsilon \rangle =
\langle 0 | \ \bar{b}(0) \gamma^\mu  \gamma_5 b(0) \ | \Upsilon \rangle
= 0 \\
&& \hspace{-1.5cm}
\langle 0 | \ \bar{b}(0)  \gamma^\mu  b(0) \ | \Upsilon \rangle =
f_\Upsilon M_\Upsilon \epsilon_\Upsilon^\mu(p), \hspace{0.5cm}
\langle 0 | \ \bar{b}(0)  \sigma^{\mu \nu} b(0) \ | \Upsilon \rangle =
- i f_\Upsilon \left[\ p^\mu ,\epsilon_\Upsilon^\nu(p)\ \right]
\label{g3}
\end{eqnarray}
where $f_\Upsilon$ is the decay constant, $M_\Upsilon$ is the mass,
$p=(M_\Upsilon, \vec{0})$ is the momentum and
$\epsilon_\Upsilon(p)$ is the polarization vector of $\Upsilon$
meson. Although the tensor current $\bar{b} \sigma^{\mu \nu} b$
is not present in (\ref{g2}), it inevitably appears in
calculations of the $\Upsilon(3S) \to \Phi \Phi^* \gamma$ amplitude. Thus, we need
this current matrix element parametrization as well.
Also, it is worth mentioning that $p \cdot \epsilon_\Upsilon = 0$ due
to the vector current conservation.

Non-local hadronic currents matrix elements  may be expressed, to the
leading order in $1/m_b$ expansion, in
terms of those of the local currents in the following way:
\begin{equation}
\langle 0 | \ \bar{b}(x_1) \ \Gamma \ b(x_2) \ | \Upsilon \rangle =
e^{-i(p/2) \cdot (x_1 + x_2)} \
\langle 0 | \ \bar{b}(0) \ \Gamma \ b(0) \ | \Upsilon \rangle
\label{g4}
\end{equation}
where $\Gamma$ is a product of $\gamma$-matrices. This relationship is derived
using the constituent quark approach, which assumes that within $\Upsilon$ meson
both $b$ and $\bar{b}$ are static and have a mass $M_{\Upsilon}/2$ each. Thus,
one neglects $O(\Lambda_{QCD})$ difference between $M_{\Upsilon}/2$ and $m_b$
as well as the quark-antiquark Fermi motion effects.

Let us proceed to the transition amplitudes and rates. As it is
discussed above, to the leading order the decay $\Upsilon(1S) \to
\Phi \Phi^*$ occurs as a result of an effective local four-particle
interaction. Thus, as it follows from (\ref{g1}), (\ref{g2}),  (\ref{g3}), only
operator $O_3$ contributes to this decay. Furthermore, if $\Phi$ is
a real scalar (or pseudoscalar) state, it is easy to show that
contribution of $O_3$ vanishes as well.

Calculation of $\Upsilon(1S) \to \Phi \Phi^*$ branching ratio is straightforward:
for $\Phi$ being a complex scalar state one gets
\begin{equation}
B(\Upsilon(1S) \to \Phi \Phi^*) =
\frac{\Gamma(\Upsilon(1S) \to \Phi \Phi^*)}{\Gamma_{\Upsilon(1S)}} =
\frac{C_3^2}{\Lambda_H^4} \frac{f_{\Upsilon(1S)}^2}{48 \pi \Gamma_{\Upsilon(1S)}}
\left[M_{\Upsilon(1S)}^2 - 4 m_\Phi^2 \right]^{3/2} \label{g5}
\end{equation}
where $m_\Phi$ is the DM particle mass and
$\Gamma_{\Upsilon(1S)} = (54.02 \pm 1.25) keV$
\cite{1} is the $\Upsilon(1S)$ total width.

For $\Phi$ being a self-conjugate spin-0 state, $\Phi = \Phi^*$, one has
\begin{equation}
B(\Upsilon(1S) \to \Phi \Phi) = 0 \label{g6}
\end{equation}
As it was mentioned above, this result is related to the fact that
the final DM particle pair state must be a P-wave, which is impossible due to
the Bose-Einstein symmetry of identical spin-0 particles. In what follows,
$\Gamma(\Upsilon(1S) \to \Phi \Phi)$ must
also vanish in higher orders in $1/m_b$ operator product
expansion.

Thus, provided that DM pair production is the dominant invisible channel,
the  signal for $\Upsilon(1S) \to  invisible$ decay would  imply that the light
spin-0 DM field has a complex nature. No evidence for the
$\Upsilon(1S) \to  invisible$ mode allows one to put some constraints on
the parameters of the models with light complex scalar Dark Matter, as we
illustrate in Sections~\ref{s8}~and~\ref{s5}.

Consider the other mode, $\Upsilon(3S) \to
\Phi \Phi^* \gamma$. As it was mentioned above,  this decay occurs
as result of a bi-local interaction, and as direct calculations show the
contribution of operator $O_3$ to the decay amplitude is vanishing, whereas
the other operators have, in general, a non-zero contribution. In the case of $\Phi$
being a complex scalar state, one gets
\begin{eqnarray}
\nonumber
A(\Upsilon(3S) \to \Phi \Phi^* \gamma) = A_1(\Upsilon(3S) \to \Phi \Phi^* \gamma) +
A_2(\Upsilon(3S) \to \Phi \Phi^* \gamma) + \\
+ A_4(\Upsilon(3S) \to \Phi \Phi^* \gamma)
\label{g7}
\end{eqnarray}
where
\begin{eqnarray}
&& \hspace{-1.5cm} A_1(\Upsilon(3S) \to \Phi \Phi^* \gamma) =
\frac{C_1}{\Lambda_H^2} \frac{2 e f_{\Upsilon(3S)}}{3 \omega} \
\epsilon^{* \mu}(k) \ \epsilon_{\Upsilon{(3S)}}^\nu (p) \
\left[ M \omega g_{\mu \nu} - p_\mu k_\nu \right] \label{g8} \\
&& \hspace{-1.5cm} A_2(\Upsilon(3S) \to \Phi \Phi^* \gamma)
= \frac{C_2}{\Lambda_H^2} \frac{2 e f_{\Upsilon(3S)}}{3 \omega} \
 \varepsilon_{\mu \alpha \nu \lambda} \
k^\mu \epsilon^{* \alpha}(k) \ p^\nu
\epsilon_{\Upsilon{(3S)}}^\lambda (p) \label{g9} \\
&&  \hspace{-1.5cm} A_4(\Upsilon(3S) \to \Phi \Phi^* \gamma)
= - \frac{C_4}{\Lambda_H^2} \frac{2 e f_{\Upsilon(3S)}}{3 \omega} \
i \varepsilon_{\mu \alpha \nu \lambda} \
k^\mu \epsilon^{* \alpha}(k) \ (p_2 - p_1)^\nu
\epsilon_{\Upsilon{(3S)}}^\lambda (p) \label{g10}
\end{eqnarray}
Here $k = (\omega, \vec{k})$ is the photon momentum, $\epsilon(k)$ is the
photon polarization vector, $p_1$ and $p_2$ are the momenta of the DM particle
and antiparticle respectively (by momentum conservation, $p_1 + p_2 + k = p$).

Note that there is a $\delta = \pi/2$ difference in the phases of $A_2$ and $A_4$.
Thus, these parts of $\Upsilon(3S) \to \Phi \Phi^* \gamma$ amplitude
do not interfere. It is also easy to check, after doing some algebra, that there
is no interference between $A_1$ and $A_2$ or $A_4$ as well.
This is also easy to understand:
contribution of the parity-conserving operator $O_1$ does not interfere with that
of the parity violating operators $O_2$ and $O_4$.

In what follows, the differential branching ratio for
$\Upsilon(3S) \to \Phi \Phi^* \gamma$ decay may be written in the following form:
\begin{eqnarray}
\nonumber
&& \hspace{-1.5cm} \frac{d B}{d \hat{s}}(\Upsilon(3S) \to \Phi \Phi^* \gamma) =
\frac{1}{\Gamma_{\Upsilon(3S)}}
\frac{d \Gamma}{d \hat{s}}(\Upsilon(3S) \to \Phi \Phi^* \gamma) = \\
&& \hspace{-1.5cm}  = \frac{d B_1}{d \hat{s}}(\Upsilon(3S) \to \Phi \Phi^* \gamma) +
\frac{d B_2}{d \hat{s}}(\Upsilon(3S) \to \Phi \Phi^* \gamma) +
\frac{d B_4}{d \hat{s}}(\Upsilon(3S) \to \Phi \Phi^* \gamma) \label{g11}
\end{eqnarray}
where $\Gamma_{\Upsilon(3S)} = (20.32 \pm 1.85) keV$ \cite{1} is the
$\Upsilon(3S)$ total width and
\begin{eqnarray}
&& \frac{d B_{1,2}}{d \hat{s}}(\Upsilon(3S) \to \Phi \Phi^* \gamma) =
\frac{C_{1,2}^2}{\Lambda_H^4} \frac{\alpha}{4 \pi}
\frac{f_{\Upsilon(3S)}^2 M_{\Upsilon(3S)}^3 (1 -\hat{s})}{54 \pi
\Gamma_{\Upsilon(3S)}} \sqrt{\frac{\hat{s} - 4 x_\Phi}{\hat{s}}} \label{g12} \\
&& \frac{d B_4}{d \hat{s}}(\Upsilon(3S) \to \Phi \Phi^* \gamma) =
\frac{C_4^2}{\Lambda_H^4} \frac{\alpha}{4 \pi}
\frac{f_{\Upsilon(3S)}^2 M_{\Upsilon(3S)}^3 (1 -\hat{s}^2)}{162 \pi
\Gamma_{\Upsilon(3S)}} \left(\frac{\hat{s} - 4 x_\Phi}{\hat{s}}
\right)^{3/2} \label{g13}
\end{eqnarray}
Here $\alpha = 1/137$ is the electromagnetic coupling constant,
$x_\Phi = m_\Phi^2/M_{\Upsilon(3S)}^2$ and
$\hat{s} = s/M_{\Upsilon(3S)}^2$, where
\begin{displaymath}
s = (p_1 + p_2)^2 = (p - k)^2 = M_{\Upsilon(3S)}^2 -
2 \ \omega M_{\Upsilon(3S)}
\end{displaymath}
is the missing mass squared. Note that the kinematically allowed range for
the missing mass squared is $4 m_\Phi^2 < s < M_{\Upsilon(3S)}^2$.
Subsequently, $4 x_\Phi < \hat{s} < 1$.

The {\it partially integrated} branching ratio for $\Upsilon(3S) \to \Phi \Phi^* \gamma$ decay,
\begin{equation}
B(\Upsilon(3S) \to \Phi \Phi^* \gamma)_{|s < s_{max}} =
\int_{4 x_\phi}^{\hat{s}_{max}}{d\hat{s} \ \frac{d B}{d \hat{s}}(\Upsilon(3S) \to \Phi \Phi^* \gamma)}
\label{i4}
\end{equation}
with $\hat{s}_{max} < 1$ or, equivalently,  $s_{max} <  M_{\Upsilon(3S)}^2$,
is given by
\begin{eqnarray}
\nonumber
B(\Upsilon(3S) \to \Phi \Phi^* \gamma)_{|s < s_{max}}  =
B_1(\Upsilon(3S) \to \Phi \Phi^* \gamma)_{|s < s_{max}} + \\
B_2(\Upsilon(3S) \to \Phi \Phi^* \gamma)_{|s < s_{max}} +
B_4(\Upsilon(3S) \to \Phi \Phi^* \gamma)_{|s < s_{max}} \label{i5}
\end{eqnarray}
where
\begin{eqnarray}
\nonumber
B_{1,2}(\Upsilon(3S) \to \Phi \Phi^* \gamma)_{|s < s_{max}}  = \frac{C_{1,2}^2}{\Lambda_H^4}
\frac{\alpha}{4 \pi}\frac{f_{\Upsilon(3S)}^2}{108 \pi \Gamma_{\Upsilon(3S)} M_{\Upsilon(3S)}}
\Biggl[ \Biggl(
2 M_{\Upsilon(3S)}^2 - s_{max} +  \\
\nonumber
+ 2 m_\Phi^2 \Biggr) \sqrt{s_{max} \left(s_{max} - 4 m_\phi^2\right)} - \\
 - \ 8 m_\Phi^2 \left(M_{\Upsilon(3S)}^2 - m_\Phi^2 \right)
 \ \ln{\left(\frac{\sqrt{s_{max}} + \sqrt{s_{max}
- 4 m_\Phi^2}}{2 m_\Phi}\right)} \Biggr] \label{i6} \\
\nonumber
B_4(\Upsilon(3S) \to \Phi \Phi^* \gamma)_{|s < s_{max}}  = \frac{C_4^2}{\Lambda_H^4}
\frac{\alpha}{4 \pi}\frac{f_{\Upsilon(3S)}^2}{162 \pi \Gamma_{\Upsilon(3S)} M_{\Upsilon(3S)}}
\Biggl\{ \Biggl[
M_{\Upsilon(3S)}^2 - \frac{s_{max}^2}{3 M_{\Upsilon(3S)}^2}+ \\
\nonumber
+ \Biggl(\frac{8 M_{\Upsilon(3S)}^2}{s_{max}} + \frac{7 s_{max}}{3 M_{\Upsilon(3S)}^2}
\Biggr) m_\Phi^2
- \frac{2 m_\Phi^4}
{M_{\Upsilon(3S)}^2} \Biggr] \sqrt{s_{max} \left(s_{max} -
4 m_\phi^2\right)} - \\
 - \ \frac{4 m_\Phi^2 \left(3 M_{\Upsilon(3S)}^4 + 2 m_\Phi^4 \right)}
{M_{\Upsilon(3S)}^2} \ \ln{\left(\frac{\sqrt{s_{max}} + \sqrt{s_{max}
- 4 m_\Phi^2}}{2 m_\Phi}\right)} \Biggr\} \label{i7}
\end{eqnarray}
We use the partially integrated branching ratio to confront the theoretical predictions
with experimental data. The existing experimental bounds on $\Upsilon \to \gamma +
invisible$ mode are final state missing mass dependent \cite{65,21}. In particular,
they may be very loose, or even there may be no bound, when the missing mass
is close to its upper
threshold, the $\Upsilon$ mass. Thus, imposing
a cutoff $s < s_{max}$ enables one to use existing experimental constraints on
$\Upsilon \to \gamma + invisible$ transition in the most efficient way. Also,
this way one gets rid of the missing mass range, where the emitted photon
energy is $O(\Lambda_{QCD})$ and non-perturbative QCD effects may be
of importance. The price we pay is shrinking of the WIMP mass range, where
our analysis is efficient.

Taking $s_{max} = M_{\Upsilon(3S)}^2$, one gets the total integrated
branching ratio for $\Upsilon(3S) \to \Phi \Phi^* \gamma$ decay:
\begin{eqnarray}
\nonumber
B(\Upsilon(3S) \to \Phi \Phi^* \gamma)  =
B_1(\Upsilon(3S) \to \Phi \Phi^* \gamma) +
B_2(\Upsilon(3S) \to \Phi \Phi^* \gamma) + \\
+ B_4(\Upsilon(3S) \to \Phi \Phi^* \gamma) \label{g14}
\end{eqnarray}
where
\begin{eqnarray}
\nonumber
\hspace{-1cm}
B_{1,2}(\Upsilon(3S) \to \Phi \Phi^* \gamma)  = \frac{C_{1,2}^2}{\Lambda_H^4}
\frac{\alpha}{4 \pi}\frac{f_{\Upsilon(3S)}^2}{108 \pi \Gamma_{\Upsilon(3S)}}
\Biggl[ \left(
M_{\Upsilon(3S)}^2 + 2 m_\Phi^2 \right) \sqrt{ M_{\Upsilon(3S)}^2 -
4 m_\phi^2} - \\
 - \ \frac{8 m_\Phi^2 \left(M_{\Upsilon(3S)}^2 - m_\Phi^2 \right)}
{M_{\Upsilon(3S)}} \ \ln{\left(\frac{M_{\Upsilon(3S)} + \sqrt{M_{\Upsilon(3S)}^2
- 4 m_\Phi^2}}{2 m_\Phi}\right)} \Biggr] \label{g15} \\
\nonumber
\hspace{-3cm}
B_4(\Upsilon(3S) \to \Phi \Phi^* \gamma)  = \frac{C_4^2}{\Lambda_H^4}
\frac{\alpha}{4 \pi}\frac{f_{\Upsilon(3S)}^2}{243 \pi \Gamma_{\Upsilon(3S)}}
\Biggl[ \left(
M_{\Upsilon(3S)}^2 + \frac{31}{2} m_\Phi^2 - \frac{3 m_\Phi^4}
{M_{\Upsilon(3S)}^2} \right) \sqrt{ M_{\Upsilon(3S)}^2 -
4 m_\phi^2} - \\
 - \ \frac{6 m_\Phi^2 \left(3 M_{\Upsilon(3S)}^4 + 2 m_\Phi^4 \right)}
{M_{\Upsilon(3S)}^3} \ \ln{\left(\frac{M_{\Upsilon(3S)} + \sqrt{M_{\Upsilon(3S)}^2
- 4 m_\Phi^2}}{2 m_\Phi}\right)} \Biggr] \label{g16}
\end{eqnarray}
The total integrated branching ratio may be used for theoretical studies, in particular to
specify if a given model or class of models may be tested with the existing level of
experimental accuracy.

In the case of $\Phi$ being a real scalar (or pseudoscalar) state, contribution
of $O_4$ vanishes and contributions of $O_1$ and $O_2$ to the branching ratio must be
multiplied by factor two. In this case, one can rewrite formulae
(\ref{g11})~-~(\ref{g16}) in a more simple form:
\begin{equation}
\frac{d B}{d \hat{s}}(\Upsilon(3S) \to \Phi \Phi \gamma) =
\frac{\left(C_1^2 + C_2^2 \right)}{\Lambda_H^4} \frac{\alpha}{4 \pi}
\frac{f_{\Upsilon(3S)}^2 M_{\Upsilon(3S)}^3 (1 -\hat{s})}{27 \pi
\Gamma_{\Upsilon(3S)}} \sqrt{\frac{\hat{s} - 4 x_\Phi}{\hat{s}}} \label{g17}
\end{equation}
\begin{eqnarray}
\nonumber
B(\Upsilon(3S) \to \Phi \Phi \gamma)_{|s < s_{max}}  =
\frac{\left(C_{1}^2 + C_2^2\right)}{\Lambda_H^4}
\frac{\alpha}{4 \pi}\frac{f_{\Upsilon(3S)}^2}{54 \pi \Gamma_{\Upsilon(3S)} M_{\Upsilon(3S)}}
\Biggl[ \Biggl(
2 M_{\Upsilon(3S)}^2 - s_{max} +  \\
\nonumber
+ 2 m_\Phi^2 \Biggr) \sqrt{s_{max} \left(s_{max} - 4 m_\phi^2\right)} - \\
 - \ 8 m_\Phi^2 \left(M_{\Upsilon(3S)}^2 - m_\Phi^2 \right)
 \ \ln{\left(\frac{\sqrt{s_{max}} + \sqrt{s_{max}
- 4 m_\Phi^2}}{2 m_\Phi}\right)} \Biggr] \label{i8}
\end{eqnarray}
and
\begin{eqnarray}
\nonumber
\hspace{-1.5cm}
B(\Upsilon(3S) \to \Phi \Phi \gamma)  = \frac{\left(C_1^2 + C_2^2 \right)}{\Lambda_H^4}
\frac{\alpha}{4 \pi}\frac{f_{\Upsilon(3S)}^2}{54 \pi \Gamma_{\Upsilon(3S)}}
\Biggl[ \left(
M_{\Upsilon(3S)}^2 + 2 m_\Phi^2 \right) \sqrt{ M_{\Upsilon(3S)}^2 -
4 m_\phi^2} - \\
 - \ \frac{8 m_\Phi^2 \left(M_{\Upsilon(3S)}^2 - m_\Phi^2 \right)}
{M_{\Upsilon(3S)}} \ \ln{\left(\frac{M_{\Upsilon(3S)} + \sqrt{M_{\Upsilon(3S)}^2
- 4 m_\Phi^2}}{2 m_\Phi}\right)} \Biggr] \label{g18}
\end{eqnarray}
As it was mentioned above, $\Upsilon(3S) \to \Phi \Phi \gamma$ is the only mode that
we use to test the models with self-conjugate spin-0 light Dark Matter.

\section{Neutrino Background}
\label{s7}
\renewcommand{\theequation}{\ref{s7}.\arabic{equation}}
\setcounter{equation}{0}

In this section we consider neutrino-antineutrino pair production within
$\Upsilon(1S) \to invisible$ and $\Upsilon(3S) \to \gamma + invisible$
channels. We examine possible impact of the neutrino
background on our analysis.

Neutrino background in $\Upsilon(1S) \to invisible$ and
$\Upsilon(3S) \to \gamma + invisible$ decays occurs due to $\Upsilon(1S) \to
\nu \bar{\nu}$ and $\Upsilon(3S) \to \nu \bar{\nu} \gamma $ transitions. These
transitions may occur both within the Standard Model and due to New Physics
interactions. The NP contribution is model dependent and should be examined
(upon necessity) in the framework of a particular model. Here we concentrate
on the Standard
Model contribution only. To the leading order within the SM, $b \bar{b} \to \nu
\bar{\nu}$ transition is mediated by virtual Z boson.

$\Upsilon \to \nu \bar{\nu}$ and $\Upsilon \to \nu \bar{\nu} \gamma$
transitions have been originally discussed in \cite{32}. Later on
$\Upsilon \to \nu \bar{\nu}$ has been studied in detail in
\cite{31}, both within the Standard Model and within of some of its
extensions. The SM result of ref.~\cite{31} may be rewritten
in terms of $\Upsilon(1S) \to \nu \bar{\nu}$ branching ratio as
\begin{equation}
B(\Upsilon(1S) \to \nu \bar{\nu}) =
\frac{\Gamma(\Upsilon(1S) \to \nu \bar{\nu})}{\Gamma_{\Upsilon(1S)}} =
\frac{N_{\nu} G_F^2}{48 \pi} \left(1 - \frac{4}{3} \sin^2{\theta_W}
\right)^2 \frac{f_{\Upsilon(1S)}^2M^3_{\Upsilon(1S)}}{\Gamma_{\Upsilon(1S)}} \label{g19}
\end{equation}
where $G_F$ is the Fermi coupling, $\theta_W$ is the weak mixing angle, and
$N_\nu = 3$ is the number of light non-sterile neutrinos. We use $G_F = 1.166
\times 10^{-5} GeV^{-2}$, $\sin^2\theta_W = 0.231$,
$M_{\Upsilon(1S)} = 9.45 GeV$ and $\Gamma_{\Upsilon(1S)} = (54.02 \pm 1.25) keV$
\cite{1}. The decay constant $f_{\Upsilon(1S)}$ may be extracted from the
experimental measurements of $\Upsilon(1S) \to e^+ e^-$ rate: one gets
 $f_{\Upsilon(1S)} = (0.715 \pm 0.005) GeV$ \cite{19}. Using indicated values of the
input parameters, one gets
\begin{equation}
B(\Upsilon(1S) \to \nu \bar{\nu}) = (1.03 \pm 0.04)
\times 10^{-5} \label{g20}
\end{equation}

Thus, $\Upsilon(1S) \to \nu \bar{\nu}$ decay
branching ratio is about 30 times less than the BaBaR
experimental bound on $B(\Upsilon(1S) \to invisible)$, given by Eq.~(\ref{i2}).
In what follows,
$\Upsilon(1S) \to \nu \bar{\nu}$ mode may be neglected in our
analysis.

Calculation of $\Upsilon(3S) \to \nu \bar{\nu} \gamma$
branching ratio within the
Standard Model is straightforward: it yields
\begin{equation}
\frac{dB(\Upsilon(3S) \to \nu \bar{\nu} \gamma)}{d \hat{s}} =
\frac{N_\nu G_F^2}{162 \pi} \frac{\alpha}{4 \pi} \frac{f_{\Upsilon(3S)}^2
M_{\Upsilon(3S)}^3}{\Gamma_{\Upsilon(3S)}} (1 - \hat{s}^2) \label{g22}
\end{equation}
and
\begin{equation}
B(\Upsilon(3S) \to \nu \bar{\nu} \gamma) =
\frac{N_\nu G_F^2}{243 \pi} \frac{\alpha}{4 \pi} \frac{f_{\Upsilon(3S)}^2
M_{\Upsilon(3S)}^3}{\Gamma_{\Upsilon(3S)}}  \label{g23}
\end{equation}
where $M_{\Upsilon(3S)} = 10.355 GeV$ and $\Gamma_{\Upsilon(3S)} = (20.32
\pm 1.85)keV$ \cite{1}, The decay constant  $f_{\Upsilon(3S)}$
may be found, using
\begin{equation}
\Gamma(\Upsilon(3S) \to e^+ e^-) = \frac{4 \pi \alpha^2 f_{\Upsilon(3S)}^2}
{27 M_{\Upsilon(3S)}} \label{i9}
\end{equation}
where $\Gamma(\Upsilon(3S) \to e^+ e^-)
= (0.443 \pm 0.008) keV$ \cite{1}. One gets
$f_{\Upsilon(3S)} = (0.430 \pm 0.004) GeV$ and, subsequently,
\begin{equation}
B(\Upsilon(3S) \to \nu \bar{\nu} \gamma) = (3.14^{+ 0.38}_{- 0.32})
\times 10^{-9} \label{g24}
\end{equation}
Thus, $\Upsilon(3S) \to \nu \bar{\nu} \gamma$ branching ratio is
about three orders of magnitude less than the BaBaR experimental
limit on $B(\Upsilon(3S) \to \gamma + invisible)$, given by Eq.~(\ref{i3}).
In what follows,
the effects related to $\Upsilon(3S) \to \nu \bar{\nu} \gamma$ decay
may be neglected as well.

Thus, we may neglect the neutrino background effects when confronting theoretical
predictions for $\Upsilon$ decays into invisible states with the experimental data.

\section{Model-Independent Bounds on the Wilson Coefficients}
\label{s8}
\renewcommand{\theequation}{\ref{s8}.\arabic{equation}}
\setcounter{equation}{0}

In this section we use our theoretical predictions and existing experimental data
to derive model-independent
constraints on the Wilson coefficients $C_i$, i=1,2,3,4, as functions of
DM particle mass, $m_\Phi$, and the heavy mass, $\Lambda_H$.

We start with the bounds, coming from the study of $\Upsilon(1S) \to \Phi \Phi^*$ decay.
These bounds will be used in the next section, to constrain the models with light
complex scalar Dark Matter.
As follows from Eq.~(\ref{g5}), the decay branching ratio depends on the Wilson coefficient
$C_3$, the WIMP mass $m_\Phi$, the heavy mass $\Lambda_H$, as well as on the
$\Upsilon(1S)$ mass, total width, and decay constant. The numerical values of
$M_{\Upsilon(1S)}$, $\Gamma_{\Upsilon(1S)}$ and $f_{\Upsilon(1S)}$ have been specified in
the previous section. Using these values, one gets
\begin{equation}
B(\Upsilon(1S) \to \Phi \Phi^*) = (5.3 \pm 0.2) \times 10^{-4} \ C_3^2 \ \left(
\frac{100 GeV}{\Lambda_H} \right)^4 \ \left[1 - \frac{4 m_\Phi^2}{M_{\Upsilon(1S)}^2}
\right]^{3/2} \label{i10}
\end{equation}
The uncertainty, about 4\%, in the numerical factor in the r.h.s. of (\ref{i10}) is
due to that in the values of the input parameters. Such a small uncertainty may
safely be neglected during the further analysis.

As it follows from Eqs.~(\ref{g20})~and~(\ref{i10}), $\Upsilon(1S) \to \Phi \Phi^*$ branching ratio
may be significantly greater than that of the $\Upsilon(1S) \to \nu \bar{\nu}$
transition. In what follows, $\Upsilon(1S)$ decay into a pair of DM particles may
be the dominant channel contributing to $\Upsilon(1S) \to invisible$. Yet, in
order to test this channel, the relevant experiments must be sensitive (at least)
to $B(\Upsilon(1S) \to invisible) \sim 10^{-4}$. This sensitivity has been reached
by the BaBaR experiment \cite{66}, as it follows from the bound on
$B(\Upsilon(1S) \to invisible)$, given by Eq.~(\ref{i2}). Substituting (\ref{i2})
into (\ref{i10}), one derives the following constraint on $|C_3|$ as a function of
$m_\Phi$ and $\Lambda_H$:
\begin{equation}
|C_3| < 0.75 \left(\frac{\Lambda_H}{100GeV}\right)^2
\left(1 - \frac{4 m_\Phi^2}{M_{\Upsilon(1S)}^2} \right)^{-3/4} \label{g25}
\end{equation}

\begin{figure}[t]
\hspace{-1.3cm}
\includegraphics[width=0.53\textwidth]{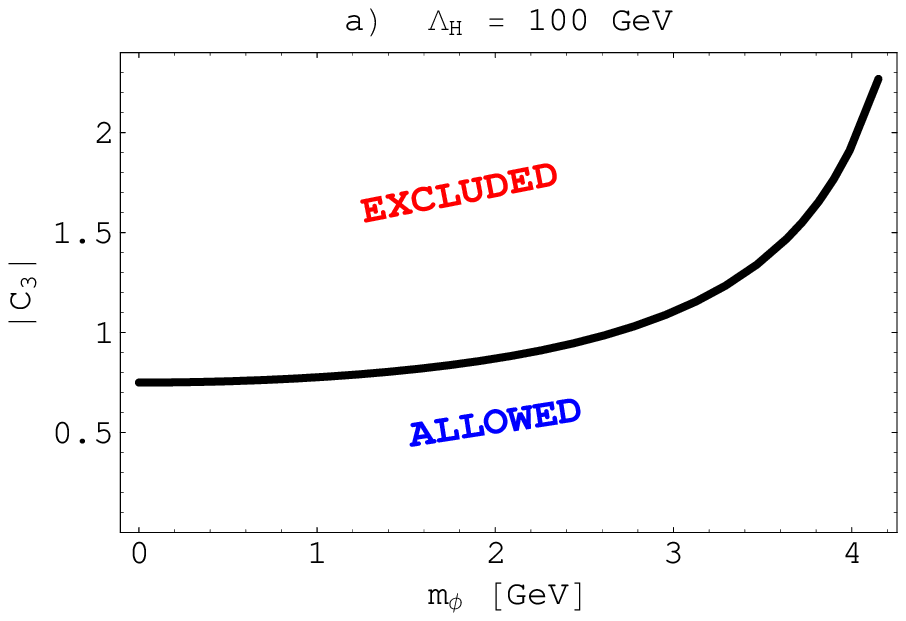} \hfill
\includegraphics[width=0.53\textwidth]{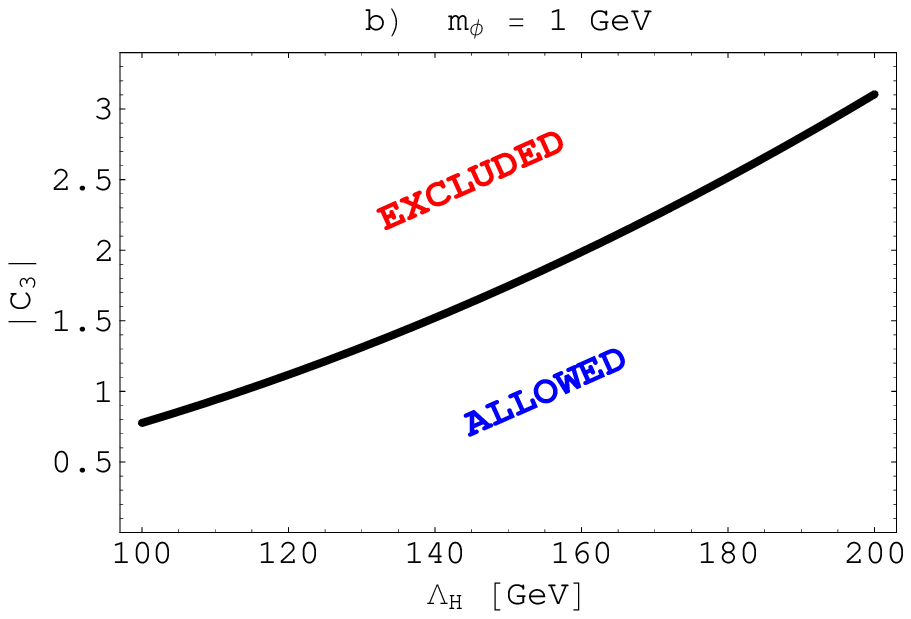}
\vspace{-0.7cm}
\caption{Upper bound on $|C_3|$ a) as a function of $m_\Phi$, for $\Lambda_H = 100~GeV$,
b) as a function of $\Lambda_H$, for $m_\Phi = 1~GeV$.}
\label{f7}
\end{figure}

The behavior of the upper bound on $|C_3|$ with the DM particle mass
is presented in Fig.~\ref{f7}(a). The bound on $|C_3|$ is almost insensitive
to the WIMP mass for
$m_\Phi < 2~GeV$; it grows rater slowly for $2~GeV < m_\Phi <  3~GeV$;
however as $m_\Phi > 3~GeV$,  the $|C_3|$ bound starts increasing rapidly,
due to the rapidly shrinking phase space. For $m_\Phi < 3~GeV$ and
 $\Lambda_H \simeq 100~GeV$,
the derived bound on $|C_3|$ may be translated into constraints on the relevant
couplings of models with a light complex spin-0 DM field. For $m_\Phi > 3~GeV$,
the experimental sensitivity to $\Upsilon(1S) \to invisible$ transition must be improved
to compensate the phase space suppression. Experimental sensitivity improvement is
also necessary for higher values of the heavy mass $\Lambda_H$: the upper bound on
$|C_3|$ grows rapidly with the heavy mass, as it can be seen from Fig.~\ref{f7}(b) and
as can be inferred from the quadratic dependence of this bound on $\Lambda_H$.

Note that within specific models, the value of $C_3$ is somehow correlated to the values
of the other Wilson coefficients, $C_1$, $C_2$ and $C_4$. In what follows, bound
(\ref{g25}) on $|C_3|$ may also lead to some constraints on $C_1$, $C_2$ and $C_4$, within
particular models with a light complex spin-0 DM field. In terms of the
branching ratios this means that experimental bound (\ref{i2}) on $B(\Upsilon(1S) \to
invisible)$ (or on $B(\Upsilon(1S) \to \Phi \Phi^*)$) may lead to {\it a phenomenological
upper bound} on $B(\Upsilon(3S) \to \Phi \Phi^* \gamma)$, within
particular models with a light complex spin-0 DM field.

At first glance, it may seem that a phenomenological upper bound on
$B(\Upsilon(3S) \to \Phi \Phi^* \gamma)$ may have little practical use,
in light of the existing BaBaR experimental constraint on $B(\Upsilon(3S)
\to \gamma + invisible)$, given by Eq.~(\ref{i3}). Yet, constraint (\ref{i3})
is derived for the emitted photon having monochromatic energy: it has been
assumed that $\Upsilon(3S) \to \gamma + invisible$ transition is mediated by
an intermediate resonant
Higgs state $A^0$ \cite{21}. Such a light Higgs state
may exist e.g. within the extensions of the Minimal Supersymmetric Standard Model
(MSSM) with an additional Higgs singlet \cite{48,24,22}.
Bounds of ref. \cite{21} on $B(\Upsilon(3S) \to \gamma + invisible)$
have been plotted as a function
of the mass of $A^0$, or equivalently, of the final state fixed missing mass.

\begin{figure}[t]
\includegraphics[width=0.9\textwidth]{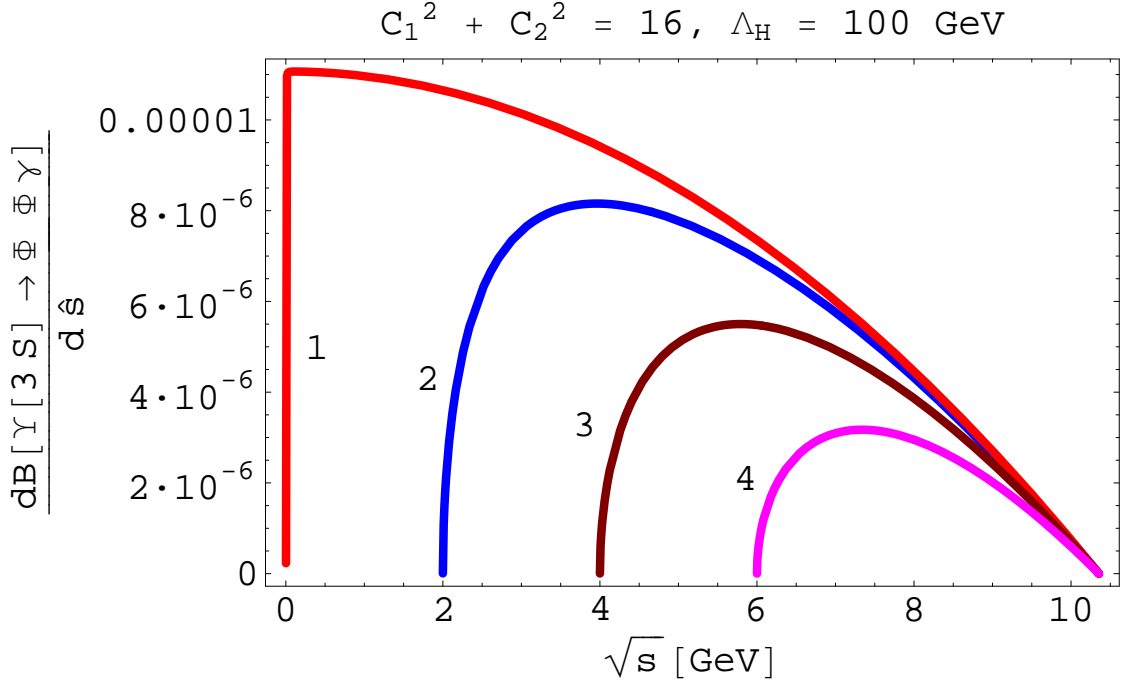}
\caption{The differential branching ratio
$dB(\Upsilon(3S) \to \Phi \Phi \gamma)/d\hat{s}$ versus the missing mass
$\sqrt{s}$ within a self-conjugate DM scenario
for $m_\Phi = 1~MeV$ (line 1), $m_\Phi = 1~GeV $ (line 2),
$m_\Phi = 2~GeV$ (line 3) and $m_\Phi = 3~GeV$ (line 4).}
\label{f2}
\end{figure}

In the case of non-resonant DM production considered here,
when the decay is mediated by heavy degrees of freedom,
the final state missing mass (or the photon energy at $\Upsilon$ rest
frame) is not fixed. Instead,  one has a broad missing mass distribution over
the entire missing mass range, $2 m_\Phi < \sqrt{s} < M_{\Upsilon(3S)}$.
For self-conjugate DM scenarios, the missing mass distribution shape
is model-independent [as it is easy to see from Eq.~(\ref{g17})] and
is depicted in Fig.~\ref{f2} for different choices of the WIMP mass.
For complex scalar DM field scenarios, the missing mass distribution analysis
depends on particular values of the Wilson coefficients [as one can
see from Eqs. (\ref{g11})-(\ref{g13})] and hence on a particular model.
It is however clear that apart from the endpoints, it is non-negligible
for the entire missing mass range as well.
In other words, the experimental analysis, performed in \cite{21}, should be extended
to the cases, when the emitted photon energy is  non-monochromatic and is
in the range $0 < \omega < M_{\Upsilon(3S)}/2 -
2 m_\Phi^2/M_{\Upsilon(3S)}$. To our knowledge, this
work is in progress now\footnote{The author is grateful to Yu. Kolomensky for this
information.}.

Note that similar problems exist with the earlier bounds
on $\Upsilon \to \gamma + invisible$, reported by CLEO \cite{65}.
Bound $B(\Upsilon(1S) \to \gamma + X) < 3 \times 10^{-5}$ is
derived assuming that $X$ is a single particle state. Thus,
the emitted photon is monochromatic again. The only existing
bound for the case of an invisible particle pair,
$B(\Upsilon(1S) \to \gamma + X \overline{X}) < 10^{-3}$, is too weak
to produce any constraints within the models with
non-resonant DM production. Due
to the factor $\alpha/(4 \pi) \approx 5.8 \times 10^{-4}$,
$B(\Upsilon \to \Phi \Phi^* \gamma)$ is always below the quoted limit.

Of course, this all does not mean that the existing experimental constrains
on $B(\Upsilon  \to \gamma + invisible)$ are totally useless, if light spin-0
DM production is mediated by heavy non-resonant degrees of freedom. Note that
BaBaR constraint (\ref{i3}) on $B(\Upsilon(3S) \to \gamma + invisible)$,
except for being plotted as a function of the final state missing mass, may
also be rewritten as a bound for a missing mass interval. One can see
from  $B(\Upsilon(3S) \to \gamma + invisible)$ versus $m_{A^0}$ plot
of ref. \cite{21} that
\begin{equation}
B(\Upsilon(3S) \to \gamma + invisible ) < 3 \times 10^{-6} \label{i13}
\end{equation}
for $\sqrt{s} \lesssim 7~GeV$ or approximately $s \lesssim M_{\Upsilon(3S)}^2/2$,
and provided that the emitted photon energy is monochromatic.

It has been discussed already that  within the complex scalar DM field scenarios,
one may derive a
phenomenological upper bound on $B(\Upsilon(3S) \to \Phi \Phi^* \gamma)$.
Comparison of this bound with (\ref{i13}) allows us to estimate
if upcoming experimental constraints on
$\Upsilon \to \gamma + invisible$, for the case of non-monochromatic photon
emission, may further improve the existing constraints on the
parameter space of a considered model. We will use this approach in the next
section.

\begin{figure}[t]
\includegraphics[width=0.9\textwidth]{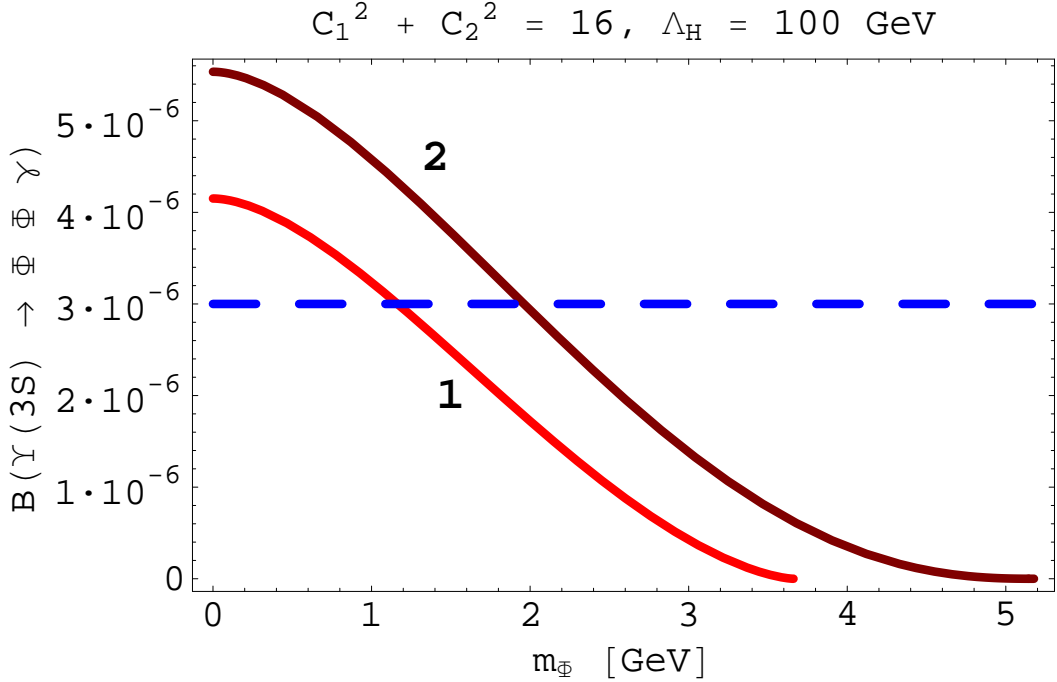}
\caption{Partially integrated  and total integrated branching ratios for
$\Upsilon(3S) \to \Phi \Phi \gamma$ decay (lines 1 and 2 respectively),
as functions of DM particle mass $m_\Phi$. The dashed line is the
experimental bound (\ref{i13}).}
\label{f3}
\end{figure}

In this section, we concentrate hereafter on self-conjugate DM scenarios
only. Recall that within these scenarios, $\Upsilon(1S) \to \Phi \Phi$ transition rate
vanishes, thus we are left with $\Upsilon(3S) \to \Phi \Phi \gamma$ channel only.
We perform the analysis of this decay channel, using the partially integrated
branching ratio, for the missing mass interval, where bound (\ref{i13}) is valid.
We remind the reader that the insertion of a
missing mass cut-off reduces significantly the WIMP mass range where an
imposed experimental bound is efficient - we illustrate this in Fig.~\ref{f3}. Yet, as it
has already been noted,
all the existing bounds on $\Upsilon \to \gamma + invisible$ have been derived
for a restricted missing mass range or for a restricted invisible particle
mass range (much below the kinematical threshold).

Using $s_{max} = M^2_{\Upsilon(3S)}/2$ and \cite{1} $M_{\Upsilon(3S)} = 10.355~GeV$,
$\Gamma_{\Upsilon(3S)} \approx 20.32~keV$,
$f_{\Upsilon(3S)} \approx 0.43~GeV$, one may rewrite formula (\ref{i8}) for
$\Upsilon(3S) \to \Phi \Phi \gamma$ partially integrated branching ratio in
the following form:
\begin{equation}
B(\Upsilon(3S) \to \Phi \Phi \gamma)_{|s < M^2_{\Upsilon(3S)}/2} = 2.6 \times 10^{-7}
\left( C_1^2 + C_2^2 \right) \left(\frac{100 GeV}{\Lambda_H} \right)^4 f(x_\Phi)
\label{i11}
\end{equation}
where
\begin{equation}
f(x_\Phi) = \left(1 + \frac{4}{3} x_\Phi \right) \sqrt{1 - 8 x_\Phi} \ - \ \frac{32}{3}
x_\Phi (1 - x_\Phi)
\ln{\left(\frac{1 + \sqrt{1 - 8 x_\Phi}}{2 \sqrt{2} \sqrt{x_\Phi}} \right)} \label{i12}
\end{equation}
Note that $0 \leq f(x_\Phi) \leq 1$, $f(x_\Phi) = 1$ if $x_\Phi = 0$, and
$f(x_\Phi) = 0$ if $x_\Phi = m_\Phi^2/M_{\Upsilon(3S)}^2 = 1/8$.

At first glance, it may seem from Eq.~(\ref{i11}) that $\Upsilon(3S) \to \Phi \Phi \gamma$
branching ratio is
far out of reach of the BaBaR experimental sensitivity, for a reasonable choice of
$C_1$ and $C_2$.  Notice, however, that within
certain models with light spin-0 Dark Matter, the Wilson coefficients $C_1$ and/or $C_2$
may be enormously large, as they contain some enhancement factors, such as the ratio
$\Lambda_H/m_b \gg 1$ -  due to the mass term in the
numerator of a heavy fermion propagator, or the Higgs vev's ratio $\tan{\beta}$
(with the latter being, say, $\sim m_t/m_b \gg 1$) - due to DM particle pair production
via exchange of a heavy non-SM Higgs degree of freedom.

These enhancement factors can make $C_1$ and/or $C_2$ to be $\gtrsim 10$
and hence
$B(\Upsilon(3S) \to \Phi \Phi \gamma)$ to be $\sim 10^{-5} - 10^{-4}$, i.e.
significantly exceeding bound (\ref{i13}) on $B(\Upsilon \to \gamma + invisible)$.
Yet, bound (\ref{i13}) assumes that the emitted photon energy is monochromatic:
rigorously speaking, one should
wait until the experimental limit on $B(\Upsilon \to \gamma + invisible)$ for
the case of non-monochromatic photon emission comes out\footnote{Also, the limit may be
derived for $\Upsilon(1S)$ state
instead of $\Upsilon(3S)$ - the author thanks Yu. Kolomensky for the discussion of this point.
The reader, however, can easily check that making the replacements
 $M_{\Upsilon(3S)} \to M_{\Upsilon(1S)}$, $f_{\Upsilon(3S)} \to f_{\Upsilon(1S)}$ and
 $\Gamma_{\Upsilon(3S)} \to \Gamma_{\Upsilon(1S)}$ in formula (\ref{i8})
 modifies our predictions by about 25\% only.}.
One may use (\ref{i13}) only to derive {\it a preliminary estimate} of
possible constraints on $\sqrt{C_1^2 + C_2^2}$ and hence
on the relevant parameters of light spin-0 self-conjugate DM models.

\begin{figure}[t]
\includegraphics[width = 0.9\textwidth]{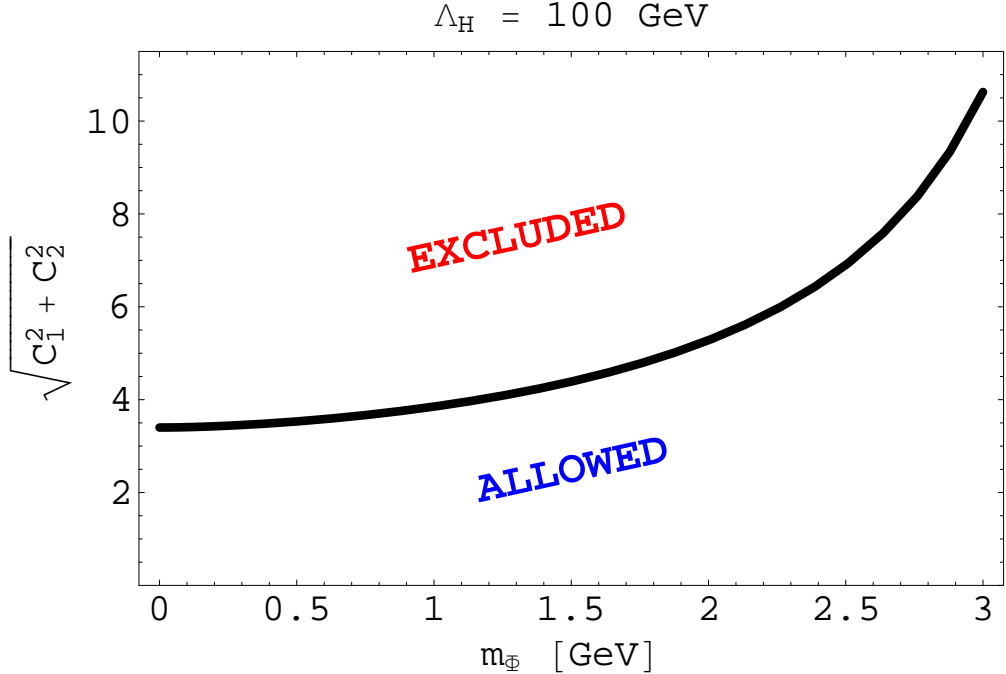}
\caption{Upper bound on $\sqrt{C_1^2 + C_2^2}$ as a function of
$m_\Phi$, for $\Lambda_H = 100~GeV$.}
\label{f4}
\end{figure}

This estimate may be presented as
\begin{equation}
\sqrt{C_1^2 + C_2^2} < 3.4 \left(\frac{\Lambda_H}{100 GeV}
\right)^2 f^{-1/2}(x_\Phi) \label{g27}
\end{equation}
It is also depicted in Fig.~\ref{f4}, as a function of $m_\Phi$, for $\Lambda_H = 100~GeV$.
Based on the discussion above, this estimate implies
rigorous constraints on the relevant
parameters of the models with light spin-0 self-conjugate Dark Matter, for $m_\Phi < 3~GeV$ .
We analyze these constraints within a particular model in Section~\ref{s4}.

\section{Complex DM Field Scenarios: Mirror Fermion Models}
\label{s5}
\renewcommand{\theequation}{\ref{s5}.\arabic{equation}}
\setcounter{equation}{0}

As it was mentioned above, the distinct feature of the scenarios with
a light complex spin-0 DM field is that $\Upsilon(1S) \to \Phi \Phi^*$ decay rate
is non-vanishing. One may use within these scenarios bound
(\ref{g25}) on $|C_3|$,
both to constrain the model parameter space, and - due to
possible correlations in the values of the Wilson coefficients - to derive
a phenomenological upper bound on the $\Upsilon(3S) \to \Phi \Phi^* \gamma$
branching ratio.

Recall that bound (\ref{g25}) on $|C_3|$ is strongest for $\Lambda_H \simeq 100~GeV$
and $m_\Phi < 3~GeV$. Yet, even for these values of the heavy and WIMP masses,
$C_3$ is still allowed
to be of the order of unity. It
seems to be very unlikely to saturate
such a (rather weak) bound, if within a full electroweak theory, $\Upsilon(1S) \to
\Phi \Phi^*$ transition is loop-induced. We may therefore restrict ourselves
by the models, where  $\Upsilon(1S) \to
\Phi \Phi^*$ decay occurs at tree level.

\begin{figure}[t]
\includegraphics[width=0.9\textwidth]{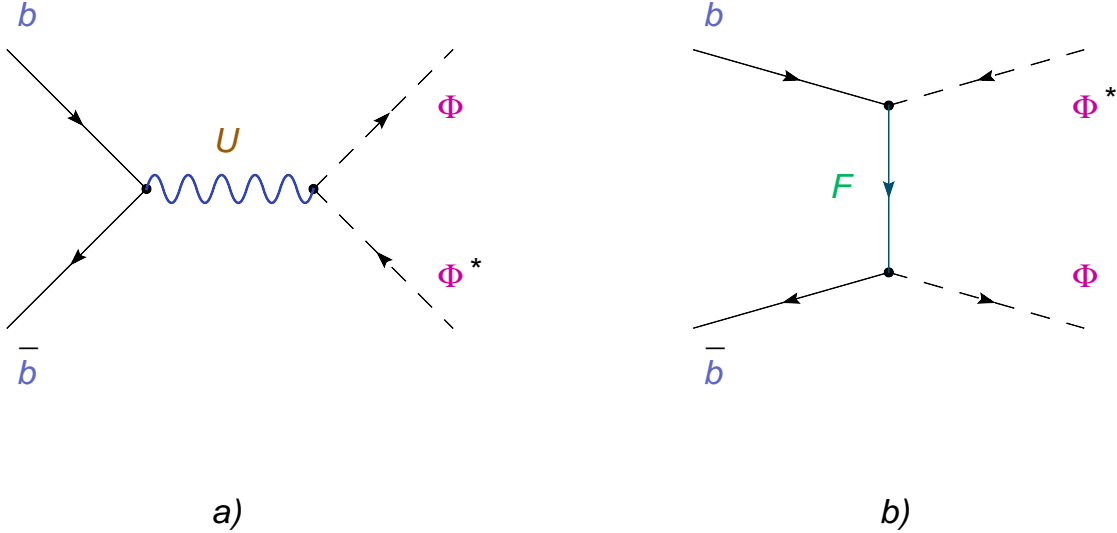}
\caption{Tree level diagrams for $\Upsilon(1S) \to \Phi \Phi^*$
decay within New Physics models. The transition occurs a) by exchange a gauge boson,
b) by exchange of a charge -1/3 fermion.}
\label{f9}
\end{figure}

Possible tree level diagrams for $\Upsilon(1S) \to \Phi \Phi^*$ transition
within New Physics models are presented in Fig.~\ref{f9}.
They involve exchange of either a neutral gauge boson or a
charge -1/3 fermion. These type of diagrams may occur e.g. within U-boson
models\footnote{Scenarios with $\Upsilon$ decaying into DM
with a large rate through the SM Z boson  may be excluded right away,
as it would imply that Dark Matter contributes to the Z boson invisible
width.}
or within mirror fermion models with light scalar Dark Matter
\cite{34,37,68,69,70}.

Models with a new (beyond the SM) neutral gauge
U-boson have been considered yet long ago, both within the supersymmetric theories
and within the other SM extensions \cite{39}~-~\cite{42}.
$\Upsilon$ decays into invisible states (with or without a photon
emission) have been studied in details in \cite{36, 37, 38},
within the scenarios with a light U-boson (with a mass less than a hundred MeV
scale). In this paper only scenarios with a heavy U-boson are of interest.
In that case, however, U-boson should be much heavier than the Z \cite{34}.
Both $\Upsilon(1S) \to \Phi \Phi^*$ and $\Upsilon(3S) \to \Phi \Phi^* \gamma$
branching ratios (inversely proportional to $m_U^4$) would
be then suppressed too much to
yield any constraints on the
parameters of U-boson models.

Thus, we concentrate here only on the models with mirror fermions,
where $\Upsilon$ meson decays into Dark Matter by exchange of
a heavy, charge -1/3 fermion, as shown in Fig.~\ref{f9}(b).
Within these models, scalar Dark Matter couples to ordinary matter
by means of Yukawa interactions, such as \cite{34,68}
\begin{equation}
- {\cal L} = \Phi \ (\lambda_{b_L} \ \overline{F}_{b_R} \ b_L \ + \ \lambda_{b_R} \
\overline{F}_{b_L} \ b_R) \ + \
 h. c. \ + \ ... \label{c1}
\end{equation}
where $F_b$ is a charge -1/3 colored mirror fermion. Unlike ordinary quarks, however,
the right-handed component of $F_b$ transforms as a
member of a weak isospin gauge doublet, while the left-handed component of
$F_b$ transforms as a singlet. In other words, $F_b$ appears
to be {\it a mirror
counterpart} of b-quark. $F_b$ and other mirror fermions do not mix with
ordinary quarks and leptons, as they - along with the DM particle - are
odd under so-called M-parity \cite{34}, or
mirror parity transformation,
whereas ordinary matter is M-even.
The scalar DM particle is the lightest M-odd particle of the theory.

Mirror fermion scenario may be realized \cite{68,69,70}
within the MSSM with gauge-mediated supersymmetry breaking and the DM particle
in the hidden sector. It has been argued \cite{68} that the thermal relic density
constraint may be satisfied irrespectively of the DM particle mass and, in
particular, within the light DM scenario (the "WIMPless miracle"). As for the
mirror fermions, they serve as connectors between the hidden and the
MSSM sectors. For the sake of simplicity, it is assumed that each connector
couples to one quark (or lepton), but other scenarios are in principle possible,
where each mirror fermion has multiple
couplings with  the SM fermions, or each quark or lepton couples to multiple
mirror fermions. The masses of mirror fermions are expected to be of the order
of electroweak braking scale or heavier\footnote{Unlike \cite{68,69,44},
we do not apply
the bound on the fourth generation quark mass, $m_{d_4} > 258~GeV$ \cite{45}.
In our opinion, this bound is irrelevant for the mirror fermion models.
Indeed, constraints coming from the annihilation channel,
$d_4 \bar{d}_4 \to q \bar{q} W W$,
are invalid here: mirror fermions may annihilate to a DM particles pair (through
interactions like that in (\ref{c1})) and hence escape detection. The other channel
in use, $d_4 \to c W$, is also invalid, due to the mirror symmetry of the model.},
i.e. $M_F \gtrsim 100~GeV$.

The tree level matching between the full and effective theories yields
\begin{eqnarray}
\nonumber
&& C_1 = - \left[ \left( \frac{M_{F_b}}{m_b^*} \right)
\frac{Re\left(\lambda_{b_R} \lambda_{b_L}^* \right)}
{2} + \frac{\left|\lambda_{b_R}\right|^2 +
\left|\lambda_{b_L}\right|^2}{4} \ \right],
\hspace{0.5cm}
C_2 = - \left(\frac{M_{F_b}}{m_b^*} \right)
\frac{Im\left(\lambda_{b_R} \lambda_{b_L}^* \right)}{2} \\
&& C_3 = - \ \frac{\left|\lambda_{b_R}\right|^2 +
\left|\lambda_{b_L}\right|^2}{4} \ , \hspace{0.5cm}
C_4 = - \ \frac{\left|\lambda_{b_R}\right|^2 -
\left|\lambda_{b_L}\right|^2}{4}, \hspace{0.5cm} \Lambda_H = M_{F_b} \label{c2}
\end{eqnarray}
where $m_b^*$ is the $\overline{MS}$ bottom mass evaluated at the matching
scale. The natural choice of the matching scale is the heavy mass,
$\mu_H = \Lambda_H$, or, according (\ref{c2}), $\mu_H = M_{F_b}$. The
bottom mass evolution with the scale is known up to four loops \cite{18},
yet to the leading-order approximation used here, it is given by
\begin{equation}
m_b^* = m_b(M_{F_b}) =
m_b(m_b) \left(\frac{\alpha_s(M_{F_b})}{\alpha_s(m_b)} \right)^{12/23}
\label{c3}
\end{equation}
where $m_b(m_b) = (4.20 \pm 0.17) GeV$ \cite{1}.

Running of $m_b$ with the scale is the only $O(1)$ QCD effect, relevant for our
analysis. As it has been already mentioned above, the Wilson coefficients $C_i$ are
{\it low-energy renormalization scale-independent}. That is to say, equations
(\ref{c2}) hold also at any scale $\mu < M_{F_b}$, including the decay scale.

We begin with using (\ref{c2}) to transform bound $(\ref{g25})$ on $|C_3|$,
as a function of $m_\Phi$ and $\Lambda_H$,
into that on the couplings $\lambda_{b_L}$ and $\lambda_{b_R}$, as functions of
$m_\Phi$ and $M_{F_b}$. This bound, in general, depends on possible correlations
in the values of $\lambda_{b_L}$ and $\lambda_{b_R}$. Here the following two
scenarios are considered:
\begin{itemize}
\item{the chiral scenario, when one of these coupling vanishes, e.g. $\lambda_{b_R} = 0$;}
\item{non-chiral scenario, with $\lambda_{b_L} = \lambda_{b_R} = \lambda_b$.}
\end{itemize}

Within the chiral scenario, the use of (\ref{g25}) and (\ref{c2}) yields
\begin{equation}
\left|\lambda_{b_L}\right| <
1.73 \left(\frac{M_{F_b}}{100 GeV} \right) \left( 1 -
\frac{4 m_\Phi^2}{M_{\Upsilon(1S)}^2} \right)^{-3/8} \label{c4}
\end{equation}
To the best of our
knowledge, bound on the coupling $\lambda_{b_L}$ (equivalently on $\lambda_{b_R}$, if
we choose instead $\lambda_{b_L} = 0$) is derived for the first time.  Dark Matter direct search
experiments, based on DM scattering off nuclei, are sensitive to the light quark - mirror fermion
interaction couplings only \cite{68,69}. So is the dominant contribution to DM
 annihilation processes.  The decays $B \to K + invisible$ or $B_s \to invisible$ would
inevitably depend both on $\lambda_{b_{L,R}}$ and on $\lambda_{s_{L,R}}$, but not
on $\lambda_{b_{L,R}}$ alone. Other quarkonium, $\chi_{b0}$, invisible decay modes still need
improvement of experimental sensitivity \cite{44}.

At first glance, bound (\ref{c4}) on $\lambda_{b_L}$ is weak. Furthermore, it is essential
only for a restricted range of the mirror fermion mass: it becomes weaker than
the perturbativity
limit, $\lambda_{b_L} < \sqrt{4 \pi}$,
if $M_{F_b} \gtrsim 200~GeV$. Nevertheless, the use of (\ref{c4}) may lead to
{\it phenomenological upper bounds} on some of bottomonium decay channels, and in
particular, on $\Upsilon(3S) \to \Phi \Phi^* \gamma$.

Indeed, for the chiral scenario, matching conditions (\ref{c2}) may be rewritten in
the following
form:
\begin{equation}
C_1 = C_3 = - C_4 = - \frac{|\lambda_{b_L}|^2}{4}, \hspace{0.5cm} C_2 = 0 \label{c5}
\end{equation}
Thus, bound (\ref{c4}) on $|\lambda_{b_L}|$
may be transformed into the constraints on the Wilson coefficients
$C_1$ and $C_4$. Then, using formulae (\ref{g14})~-~(\ref{g16}), for $C_2=0$ and the
other parameters values specified in Sections~2~and~3, and applying
the constraints on $C_1$ and $C_4$, one derives
an upper bound on $\Upsilon(3S) \to \Phi \Phi^* \gamma$ (total) branching ratio, as a
function of DM particle mass, $m_\Phi$.

\begin{figure}[t]
\includegraphics[width=0.9\textwidth]{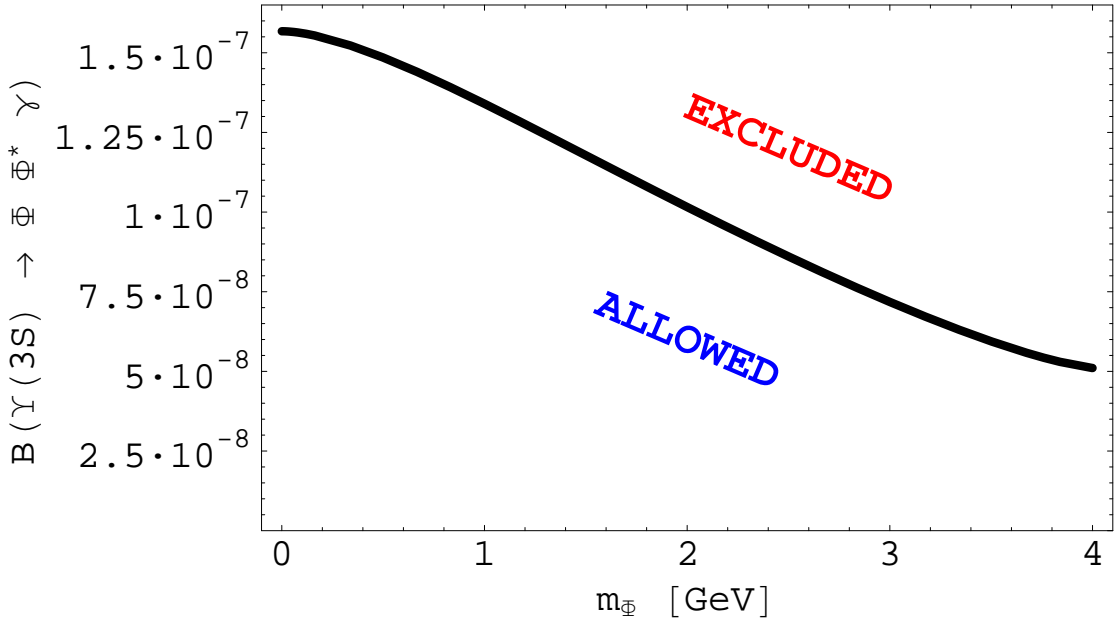}
\caption{Phenomenological upper bound on $B(\Upsilon(3S) \to \Phi \Phi^* \gamma)$ branching
ratio within the chiral scenario.}
\label{f8}
\end{figure}

We present this bound in Fig.~\ref{f8}. It may also be rewritten as
$B(\Upsilon(3S) \to \Phi \Phi^* \gamma) < 1.57 \times 10^{-7}$, with the limit
being saturated, as $m_\Phi \to 0$. Certainly, within the chiral scenario,
$B(\Upsilon(3S) \to \Phi \Phi^* \gamma)$ is far away of the present
experimental sensitivity. This result is not surprising: it has been anticipated
in the previous section, for the models with no enhancement factors in
the Wilson coefficients. The use of the derived constraint (\ref{c4})
on $|\lambda_{b_L}|$
enables one to formulate this quantitatively, by imposing a
well defined limit on the $\Upsilon(3S) \to \Phi \Phi^* \gamma$ branching
ratio.

\begin{figure}
\includegraphics[width=0.9\textwidth]{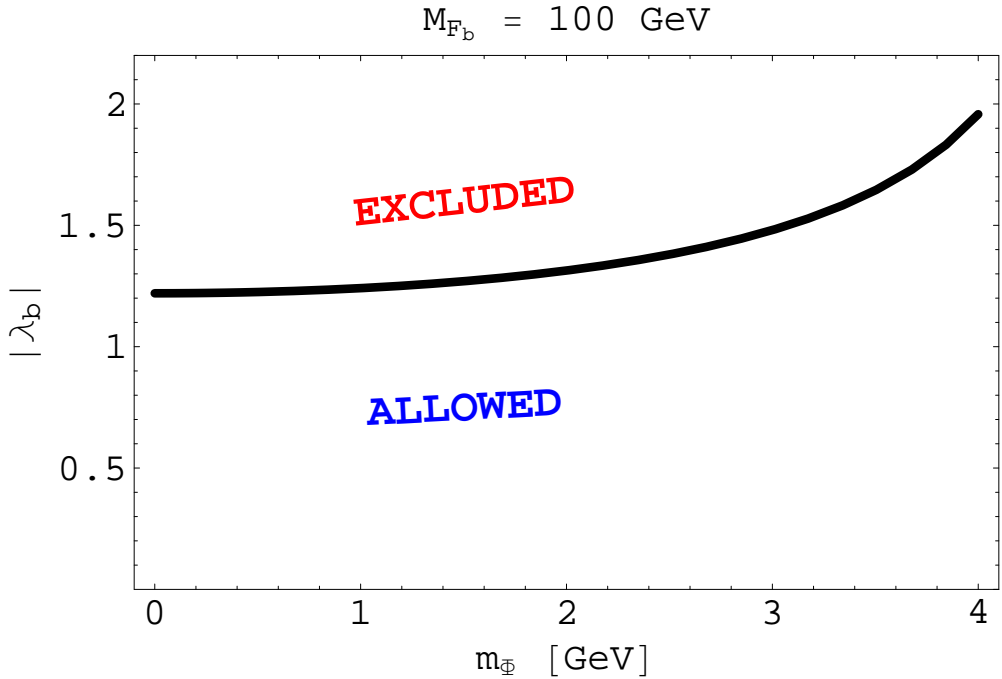}
\caption{Upper bound on $|\lambda_b|$ as a function of $m_\Phi$, for $M_{F_b} = 100~GeV$.}
\label{f10}
\end{figure}

It is worth noting that a signal for the
$\Upsilon(3S) \to \gamma + invisible$ transition, above the quoted limit on
$B(\Upsilon(3S) \to \Phi \Phi^* \gamma)$, would
rule out the mirror fermion models with light complex spin-0 Dark Matter and
chiral mirror couplings. The following statement is also true: such a signal
would imply that light spin-0 DM field is self-conjugate within the mirror fermion
models with chiral mirror couplings.
This result is not surprising as well: it has been noted previously
that study of $\Upsilon \to \Phi \Phi^*$ transition (combined with that of
$\Upsilon \to \Phi \Phi^* \gamma)$ represents an opportunity to test, whether
Dark Matter, if being light and a spin-0 state, is self-conjugate or
complex.

Within the non-chiral scenario, $\lambda_{b_L} = \lambda_{b_R} = \lambda_b$, noting
that $C_3 = - |\lambda_b|^2/2$, we may rewrite bound (\ref{g25}) on $|C_3|$ in
the following form:
\begin{equation}
\left|\lambda_{b}\right| <
1.22 \left(\frac{M_{F_b}}{100 GeV} \right) \left( 1 -
\frac{4 m_\Phi^2}{M_{\Upsilon(1S)}^2} \right)^{-3/8} \label{c6}
\end{equation}
We present this bound in Fig.~\ref{f10}, as a function of $m_\Phi$, for
$M_{F_b} = 100~GeV$. The derived constraint on $|\lambda_b|$ changes
rather slowly with the WIMP mass, if $m_\Phi < 3~GeV$. To simplify
the analysis, one may use, within a crude approximation,
$|\lambda_b| < 1.22$ for this range of the WIMP mass and $M_{F_b} \simeq 100~GeV$.
The behavior
of bound (\ref{c6}) with the mirror fermion mass is rather trivial
($\max[|\lambda_b|]$ grows linearly with $M_{F_b}$). We note here only
that (\ref{c6}) should be replaced by the perturbativity limit,
$\lambda_b < \sqrt{4 \pi}$, if
$M_{F_b} \gtrsim 300~GeV$.

Generally speaking, it is expected that ordinary-to-mirror fermion couplings
be significantly less than one \cite{34,68}. If these
couplings are of the same order for all three quark generations, one expects
$\lambda_b \sim \lambda_{u,d} \sim
0.1 \left(M_{F_b}/100 GeV \right)$ \cite{34} from the DM relic abundance
condition, $\Omega_{DM} h^2 \sim 0.11$. Otherwise, if a hierarchy in
the couplings exists, one may
argue that this expectation does not hold for
$\lambda_b$:  its value is not affected by the Dark Matter direct search
designated experiments, nor does it have essential impact on DM annihilation and, hence,
relic abundance. Furthermore, having a third generation
Yukawa coupling to be of order unity is not unusual.
However, bound (\ref{c6}) on $|\lambda_b|$ may be significantly improved
due to possible constraints coming from the study
of $\Upsilon \to \Phi \Phi^* \gamma$ transition.

Indeed, within the non-chiral scenario,
matching conditions (\ref{c2}) may be rewritten for the Wilson coefficients
$C_1$, $C_2$ and $C_4$ as
\begin{equation}
C_1 = \left(\frac{M_{F_b}}{m_b^*} \right) \frac{|\lambda_b|^2}{2},
\hspace{0.5cm} C_2 = C_4 = 0  \label{c7}
\end{equation}
Because of the enhancement factor $M_{F_b}/m_b^*$, $C_1$ may be enormously large:
as it has been discussed in Section~\ref{s8}, $\Upsilon \to \Phi \Phi^* \gamma$
branching ratio is then within the reach of the present experimental sensitivity.
Furthermore, due to this enhancement factor,
$B(\Upsilon(3S) \to \Phi \Phi^* \gamma) \propto 1/M_{F_b}^2$ roughly\footnote{Slight deviation
from this rule occurs due to running of the bottom mass with the heavy mass scale.}, unlike
$B(\Upsilon(1S) \to \Phi \Phi^*) \propto 1/M_{F_b}^4$. Thus, within  the non-chiral scenario,
$\Upsilon(3S) \to \Phi \Phi^* \gamma$ mode may be sensitive to a wider range of the
mirror fermion mass than $\Upsilon(1S) \to \Phi \Phi^*$.

For the numerical analysis, we use $|\lambda_b| < 1.22$. As it is mentioned above,
this bound may be used in a crude approximation for $m_\Phi < 3~GeV$ and
$M_{F_b} = 100~GeV$. In order to highlight the behavior of
$B(\Upsilon(3S) \to \Phi \Phi^* \gamma)$ with the mirror fermion mass properly,
we keep using
$|\lambda_b| < 1.22$ for $M_{F_b} > 100~GeV$ as well (even though the constraint on
$|\lambda_b|$ is significantly weaker then). This yields
$B(\Upsilon(3S) \to \Phi \Phi^* \gamma) < 1.23 \times 10^{-4}$,
$B(\Upsilon(3S) \to \Phi \Phi^* \gamma) < 3.4 \times 10^{-5}$ and
$B(\Upsilon(3S) \to \Phi \Phi^* \gamma) < 0.93 \times 10^{-5}$, for $M_{F_b} = 100~GeV$,
$M_{F_b} = 200~GeV$ and $M_{F_b} = 400~GeV$ respectively, with the limits being saturated,
as $m_\Phi \to 0$. We compare these limits to bound (\ref{i13}) on $B(\Upsilon(3S)
\to \gamma + invisible)$. Recall that bound (\ref{i13}) is derived for the emitted
photon having monochromatic energy, thus it may be invalid for the scenarios where
light DM is produced due to exchange of heavy non-resonant degrees of freedom. Yet,
comparison of the above derived limits on $B(\Upsilon(3S) \to \Phi \Phi^* \gamma)$
to (\ref{i13}) allows us to infer that upcoming BaBaR constraint on $\Upsilon
\to \gamma + invisible$, for the case of non-monochromatic photon emission, may
essentially improve the existing bound on $|\lambda_b|$.

If this constraint is of the same order as (\ref{i13}), one would get
$|\lambda_b| \lesssim 0.5$, $|\lambda_b| \lesssim 0.65$ and
$|\lambda_b| \lesssim 0.9$, for sufficiently low WIMP mass and
for $M_{F_b} = 100~GeV$, $M_{F_b} = 200~GeV$ and
$M_{F_b} = 400~GeV$ respectively. The upcoming experimental data, however, may
be not so optimistic, they may lead to a much weaker experimental limit than (\ref{i13}).
In either case,  we want to emphasize again that we derive here
 {\it preliminary estimates only} of possible improvements of the existing bound on
 $|\lambda_b|$. More rigorously, regardless of any expectation,
 one must presently use the existing  bound on $|\lambda_b|$,
 given by Eq.~(\ref{c6}).

To summarize our discussion of the mirror fermion models, we note that
study of $\Upsilon(1S)$ decay into a spin-0 DM particles pair,
$\Upsilon(1S) \to \Phi \Phi^*$, leads -
for the first time - to the bounds on the couplings of b-quark interaction
with its mirror counterpart. Within the non-chiral scenario, these bounds may be somewhat
improved
by upcoming constraints on the $\Upsilon \to \gamma + invisible$ mode
for a non-monochromatic
photon emission. Recall also that we assumed throughout this section that light spin-0
DM field has complex nature. Otherwise, within the self-conjugate DM scenario, no
constraints on $\lambda_b$ couplings, apart from preliminary estimates,
may presently be derived.

\section{Self-Conjugate DM Scenario: Dark Matter Model with two Higgs Doublets (2HDM)}
\label{s4}
\renewcommand{\theequation}{\ref{s4}.\arabic{equation}}
\setcounter{equation}{0}

In this section we will discuss the models, where interaction of
(self-conjugate) spin-0 Dark Matter with the ordinary matter is mediated by heavy
Higgs degrees of
freedom. These models are known to be the most economical SM extensions,
containing Dark Matter: they are created by extending (if necessary) the SM Higgs sector
and embedding a scalar DM particle into theory, by adding a rather small number of unknown
parameters.

The simplest model of this type is the Minimal Scalar Dark Matter Model \cite{10, 11, 2}.
It has the same particle content as the SM, plus a gauge singlet real scalar field
$\Phi$, odd under $Z_2$ discrete symmetry ($\Phi \to - \Phi$) and
coupled to the SM particles through the exchange of Higgs boson. This model has been
widely
studied in the literature \cite{2,15,16}, \cite{10}-\cite{9}. Presently, the Minimal
Scalar DM Model has a very restrained parameter space \cite{2,14,15,16}, if
the DM particle is chosen to have a GeV or smaller mass. Also, the
$\Upsilon$ decay channels considered
here are not sensitive to the parameter space of
this model.
Indeed, as it has been shown in Section~\ref{s8},
for $B(\Upsilon(3S) \to
\Phi \Phi \gamma)$ to be within the reach of the present experimental sensitivity,
the Wilson coefficients $C_1$ and $C_2$ must contain some enhancement factors, such
as heavy-to-light mass ratio or a large Higgs vev's ratio. None of these enhancement
factors may be generated in the Minimal Scalar Dark Matter Model, where DM particle
production
occurs solely via the SM Higgs boson exchange. Recall also that the other decay,
$\Upsilon \to \Phi \Phi^*$, rate vanishes, when $\Phi=\Phi^*$.
Instead, the Minimal Scalar DM
Model is well tested by $B \to K + invisible$ mode, if assuming that
DM particle has
a GeV or smaller mass. Study of this mode
leads to rigorous bounds on the model parameter space  \cite{16,15}.

In this section we consider the simplest extension of the Minimal Scalar DM Model,
the two-Higgs doublet model (2HDM) with
a gauge singlet real scalar DM particle.
The DM interaction part of Lagrangian, relevant for our analysis, may be written as
\cite{16}
\begin{equation}
-{\cal L} = \frac{m_0^2}{2} \Phi^2 + \lambda_1 \Phi^2 |H_1|^2 +
\lambda_2 \Phi^2 |H_2|^2 + \lambda_3 \Phi^2 \left(H_1 H_2 + h.c \right) \label{t1}
\end{equation}
where $\Phi$ is a $Z_2$ odd real scalar DM field and
\begin{displaymath}
H_1 = \left( \begin{array}{c}
H_1^+ \\ H_1^0 \end{array} \right), \hspace{0.5cm}
H_2 = \left( \begin{array}{c}
H_2^0 \\ H_2^- \end{array} \right), \hspace{0.5cm}
H_1 H_2 = H_1^0 H_2^0 - H_1^+ H_2^-
\end{displaymath}
Following Refs.~\cite{16,9}, we consider type II version of 2HDM,
where $H_1$ generates masses of down-type quarks and
charged leptons, whereas $H_2$ generates masses of up-type
quarks\footnote{The used definition of $H_1$ and $H_2$ corresponds
to the following notations for the Yukawa interactions:
$-{\cal L}_Y = \sum_f{\left[h_{\ell_f} \bar{L}_f H_1 \ell_f
+ h_{d_f} \bar{Q}_f H_1 d_f +
h_{u_f} \bar{Q}_f H_2 u_f \right]}$.}.
The Higgs vev's, $v_1$, $v_2$, are constrained by the condition $v_1^2 +
v_2^2 = v^2= (246 GeV)^2$. The Higgs vev's ratio, $\tan{\beta} \equiv v_2/v_1$,
is a free parameter of the theory. We assume here that $\tan{\beta} \gg 1$.
As discussed before, $B(\Upsilon(3S) \to \Phi \Phi \gamma)$ is then enhanced
and may be within the reach of the BaBaR experimental sensitivity. Large
$\tan{\beta}$ scenarios are also known to be much less restrained than
the Minimal Scalar DM Model. One may avoid tension with satisfying
the DM relic density constraint. Another point worth mentioning is
that the fine-tuning of parameters needed to generate a GeV or smaller
DM mass  is, in general, significantly weaker \cite{16} than that within the
Minimal Scalar DM Model.

After eliminating Goldstone modes, one may express $H_1$ and $H_2$ in
terms of two CP-even, one CP-odd and two complex charged
mass eigenstates -  $h^0$, $H^0$, $A^0$, $H^{\pm}$ respectively
\cite{25}, \cite{9}:
\begin{eqnarray}
\nonumber
H_1^0 = \frac{1}{\sqrt{2}} \left( v_1 + H^0 \cos{\xi} - h^0 \sin{\xi} +
i A^0 \sin{\beta} \right),
\hspace{0.5cm}
H_1^+ = H^+ \sin{\beta} \\
H_2^0 = \frac{1}{\sqrt{2}} \left(v_2  + h^0 \cos{\xi} + H^0 \sin{\xi} +
i A^0 \cos{\beta}\right), \hspace{0.5cm}
H_2^- = H^- \cos{\beta}  \label{t5}
\end{eqnarray}
Here $h^0$ and $H^0$ are the lightest and heaviest CP-even states respectively,
and $\xi$ is the CP-even mass-matrix diagonalization  angle.

Using Eqs.~(\ref{t5}), one may
rewrite (\ref{t1}) in the following form:
\begin{equation}
 - {\cal L} \ = \ \frac{m_\Phi^2}{2} \Phi^2 \ + \ \lambda_{h^0 \Phi} \ v \ \Phi^2 h^0 \ + \
 \lambda_{H^0 \Phi} \ v \ \Phi^2 H^0 \ + \ \ldots \label{t10}
\end{equation}
where
\begin{equation}
m_\Phi^2 = m_0^2 + \left(\lambda_1 \cos^2{\beta}  + \lambda_2 \sin^2{\beta} +
2 \lambda_3 \sin{\beta} \cos{\beta} \right) v^2 \label{t2}
\end{equation}
is the DM particle mass,
\begin{eqnarray}
&& \lambda_{h^0 \Phi} = - \lambda_1 \sin{\xi} \cos{\beta} +
\lambda_2 \cos{\xi} \sin{\beta} + \lambda_3 \cos(\xi + \beta), \label{t11}
\\ 
&& \lambda_{H^0 \Phi} =  \lambda_1 \cos{\xi} \cos{\beta} +
\lambda_2 \sin{\xi} \sin{\beta} + \lambda_3 \sin(\xi + \beta), \label{t12}
\end{eqnarray}
and the ellipsis in (\ref{t10})
stands for the dropped quartic interaction terms. To the leading order
in the perturbation theory,
$\Upsilon(3S) \to \Phi \Phi \gamma$ transition occurs at tree level by
exchange of a single Higgs boson. Thus, to the leading-order approximation,
it is sufficient to consider only the cubic interaction terms in (\ref{t10}).
Then, as it follows from (\ref{t10}), only the CP-even Higgs states are relevant
for our analysis.

The b-quark Yukawa interaction terms may be written in the following form
\cite{25,9,1}:
\begin{equation}
-{\cal L}_{Y} = - \frac{m_b}{v} \frac{\sin{\xi}}{\cos{\beta}} \ \bar{b} \ h^0 \ b \ + \
\frac{m_b}{v} \frac{\cos{\xi}}{\cos{\beta}} \ \bar{b} \ H^0 \ b + \ldots \label{t13}
\end{equation}
where we write down explicitly only the b-quark interactions
with the CP-even Higgs bosons.

Our strategy is the following now. We use (\ref{t10}), (\ref{t11})~-~(\ref{t13}), to
derive the matching conditions for the Wilson coefficients $C_1$ and $C_2$. Then,
we transform bound (\ref{g27}) on
$\sqrt{C_1^2 + C_2^2}$ into that on the relevant couplings of the model, depending
on the DM particle mass, Higgs mass and $\tan{\beta}$. We have to recall, however,
that this bound may serve only as {\it a preliminary estimate} of possible
constraints on the parameter space of the model.

In the limit of $\tan{\beta} \gg 1$, the CP-even mixing angle $\xi$ has two
possible solutions \cite{25}:
\begin{itemize}
\item[(a)]{$\xi \approx \pi/2 - \beta$}
\item[(b)]{$\xi \approx - \beta$}
\end{itemize}

In the former case (case a), the lightest CP-even Higgs boson, $h^0$,
is Standard Model-like:
its phenomenology is similar to that of the SM Higgs boson and the experimental bound
on its mass is close to the SM limit\footnote{Also, if
the SM Higgs decays predominantly invisibly,
the SM lower experimental bound is distorted by a few GeV only \cite{29,30}.}
(see \cite{1} and references therein).

In the latter case (case~b), $h^0$ is "New-Physics (NP) like": its phenomenology differs
drastically from that of
the SM Higgs boson \cite{25}.
In particular, $m_{h^0}$ may be much
below the Standard Model experimental limit: according the existing experimental
data \cite{27,28}, $m_{h^0} > 55~GeV$ or $m_{h^0} < 1~GeV$ in the general type~II
2HDM.

As it was mentioned above, the light Higgs scenario is beyond the scope of the
present paper, thus we assume here that $m_{h^0} > 55~GeV$.  Notice, however, that this
bound is derived, provided that no invisible Higgs decay mode exists. On the other hand,
if the NP-like Higgs invisible decay mode is dominant, it may escape
detection. No bound on $m_{h^0}$, to our best knowledge, exists in that case.

Within the considered model with light scalar Dark Matter,
analysis of the $\Upsilon \to \Phi \Phi \gamma$ mode may restrict
the scenarios with an invisibly
decaying lightest Higgs boson by putting severe constraints on the $h^0 \Phi \Phi$
interaction coupling,
$\lambda_{h^0 \Phi}$, given by Eq.~(\ref{t11}).
If the lightest CP-even Higgs boson is NP-like
($\xi \approx - \beta$, case b), one may also rewrite (\ref{t11}) as
\begin{equation}
\lambda_{h^0 \Phi} \approx \lambda_3 + \left(\lambda_1 +
\lambda_2 \right) \cos{\beta}  \label{l9}
\end{equation}
The last term in the r.h.s. of (\ref{l9}),
although being suppressed by a factor of
$\cos{\beta} \approx 1/\tan{\beta}$, must be retained because of
possible hierarchy in the values of $\lambda_3$ and $\lambda_1$ or $\lambda_2$.
Scenarios with such a hierarchy may be  of importance, as
$\lambda_{h^0 \Phi}$ is constrained to be $O(1/\tan{\beta})$, if $h^0$ mass
approaches to its lower limit, $m_{h^0} = 55~GeV$.

Indeed, neglecting the heaviest SM-like CP-even Higgs
exchange contribution, the matching conditions for the Wilson coefficients have the
following form:
\begin{equation}
C_1 = - \frac{\lambda_{h^0 \Phi}}{2} \tan{\beta}, \hspace{0.5cm}
C_2 = 0, \hspace{0.5cm} \Lambda_H = m_{h^0} \label{t15}
\end{equation}
Thus, one may rewrite bound (\ref{g27}) on $\sqrt{C_1^2 + C_2^2}$ as
\begin{equation}
|\lambda_{h^0 \Phi}| <
\left(\frac{2.1}{\tan{\beta}} \right)
\left(\frac{m_{h^0}}{55 GeV} \right)^2 f^{-1/2}(x_\Phi) \label{l5}
\end{equation}
where $f(x_\Phi)$, $x_\Phi = m_\Phi^2/M_{\Upsilon(3S)}^2$, is given by Eq.~(\ref{i12}).

\begin{figure}
\includegraphics[width=0.9\textwidth]{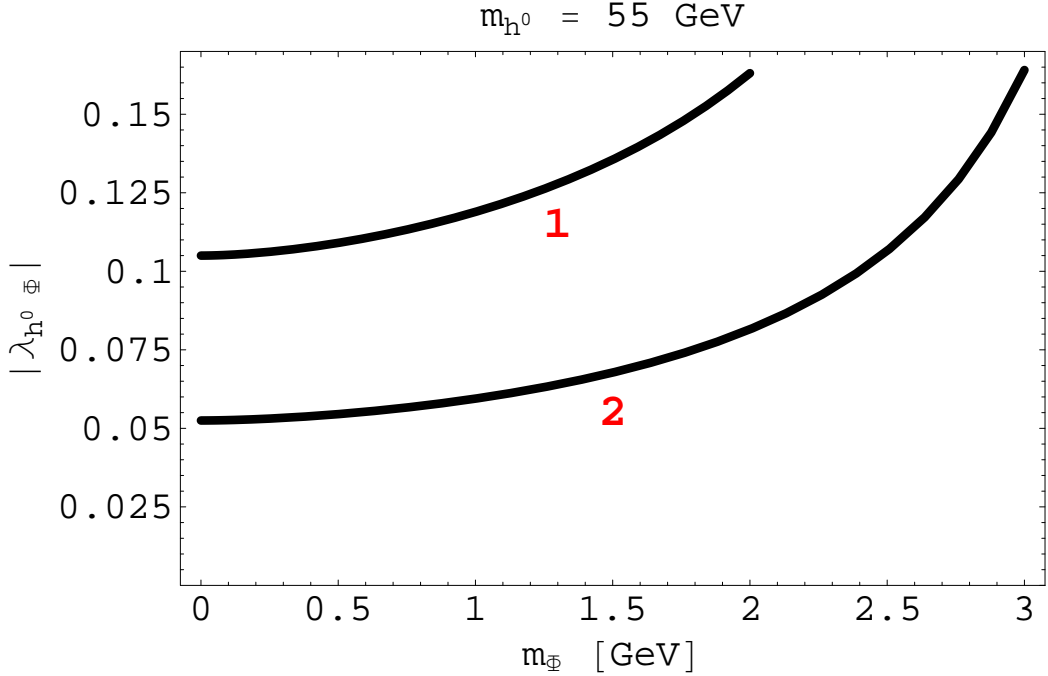}
\caption{Upper bound on $|\lambda_{h^0 \Phi}|$ as a function of $m_\Phi$, for $m_{h^0} = 55~GeV$
and $\tan{\beta} = 20$ (line 1), $\tan{\beta} = 40$ (line 2).}
\label{f6}
\end{figure}

Formula (\ref{l5}) implies a stringent upper bound on $|\lambda_{h^0 \Phi}|$ at the Higgs mass
lower threshold, $m_{h^0} = 55~GeV$. Choosing e.g.  $\tan{\beta} = 20$, one
has  $|\lambda_{h^0 \Phi}| \lesssim 0.1$ for the WIMP mass, $m_\Phi$, less than 2 GeV,
as one can see from Fig.~\ref{f6}.
For the other choice of the Higgs vev's ratio, $\tan{\beta} = 40$, the same constraint
on $|\lambda_{h^0 \Phi}|$ is derived for $m_{\Phi} = (2.5 - 3) GeV$, one also gets
$|\lambda_{h^0 \Phi}| \lesssim 0.05$,
as $m_\Phi < 1~GeV$. For such small values
of $|\lambda_{h^0 \Phi}|$, it seems to be very unlikely that $h^0$
would escape detection
and its mass be below 55 GeV.
More rigorously, however,
detailed reanalysis of the Higgs production and decay rates, including that of
$h^0 \to \Phi \Phi$,  should be performed (which is beyond the scope of the present paper).

Bound on $|\lambda_{h^0 \Phi}|$ may still be rigorous, if
the $h^0$ mass is heavier
than 55 GeV. For instance, choosing $m_{h^0} = 100~GeV$ (in which case $h^0$ is still
NP-like) and $m_\Phi < 2~GeV$, one
gets $|\lambda_{h^0 \Phi}| \lesssim 0.2$ for $\tan{\beta} = 40$, $m_\Phi < 1.5~GeV$ and
for $\tan{\beta} = m_t(m_t)/m_b(m_t) \approx 55$, $m_\Phi < 2~GeV$. Yet,
$\lambda_{h^0 \Phi}$ constraints become weaker with decreasing
$\tan{\beta}$ and increasing WIMP mass. It may be of order of the
SM weak coupling, if $\tan{\beta} \simeq 20$ and $m_\Phi \simeq 2.4~GeV$.

Thus, within the type~II 2HDM with a light spin-0 Dark Matter,
study of $\Upsilon \to \Phi \Phi \gamma$ decay channel may lead to severe constraints
on the lightest CP-even Higgs invisible decay coupling, if that Higgs is New-Physics like.

As it follows from Eq.~(\ref{l9}), bound (\ref{l5}) on $|\lambda_{h^0 \Phi}|$
may also be transformed into that on
the couplings $\lambda_1$, $\lambda_2$ and $\lambda_3$. Constraints on
$\lambda_1$ and/or $\lambda_2$ may also be derived from the study of
$B \to K + invisible $ transition \cite{16}. Those, in general, are strong enough:
in particular,
$O(1)$ values of $\lambda_1$ and $\lambda_2$ for $m_{h^0} \simeq 55~GeV$ are
ruled out, if $m_\Phi \lesssim 1.5~GeV$. In that case, $1/\tan{\beta}$ suppressed terms
in Eq.~(\ref{l9}) may safely be neglected, so that $\lambda_{h^0 \Phi} \approx
\lambda_3$. In other words, bound (\ref{l5}) on $|\lambda_{h^0 \Phi}|$ is also that
on $|\lambda_3|$, if $m_\Phi \lesssim 1.5~GeV$.

Note that due to cancellation effects in the relevant diagrams, WIMP pair production
rate in B meson decays
is insensitive to the value of $\lambda_3$ \cite{16},
and hence,
in general, to the value of $\lambda_{h^0 \Phi}$. The scenarios with $\lambda_3$
dominant, or at least non-negligible, have been thus far unconstrained. To
our best knowledge, bound on the $h^0 \Phi \Phi$ interaction coupling,
$\lambda_{h^0 \Phi}$,  is derived for the first time.
Thus, study of DM production in $\Upsilon$ decays
enables one to test the regions of the parameter space of
2HDM with scalar Dark Matter, which are
inaccessible by B meson decays with missing energy.

It may seem naively that the rigorous constraint on the
$h^0 \Phi \Phi$ interaction coupling,
$\lambda_{h^0 \Phi}$, results also to suppress the DM annihilation rate, which in its turn
may lead to scenarios with overabundant Dark Matter.
Yet, the DM annihilation rate to down-type quarks and charged
leptons is enhanced by a factor of $\tan^2{\beta}$ (due to $\tan{\beta}$ enhancement of these
fermions Yukawa interaction with the CP-even Higgs, if the one is NP-like).
Or, equivalently, using the notations of
ref. \cite{16}, the effective coupling for these annihilation processes is
\begin{displaymath}
\kappa_{h^0 \Phi} =
\lambda_{h^0 \Phi}  \left(\frac{100 GeV}{m_{h^0}}\right)^2 \tan{\beta}
\end{displaymath}
In terms of the coupling $\kappa_{h^0 \Phi}$, bound (\ref{l5}) may be
rewritten as
\begin{equation}
\kappa_{h^0 \Phi} < 6.8 \ f^{-1/2}(x_\Phi) \label{l6}
\end{equation}
Keeping in mind that $0 < f(x_\Phi) < 1$ or, equivalently,  $1 < f^{-1/2}(x_\Phi) < \infty$,
bound (\ref{l6}) on $\kappa_{h^0 \Phi}$
yields no essential constraints on DM annihilation and, hence, no tension with
satisfying DM relic abundance constraint above the Lee-Weinberg limit of the
model, $m_\Phi \gtrsim 100~MeV$ \cite{16}.
Thus, the 2HDM scenario with $\lambda_3$
dominant, or at least non-negligible, is much less restrained, than those with
$\lambda_1$ or $\lambda_2$ dominant or than Minimal Scalar DM Model \cite{16,15}.

So far it has been assumed that the lightest CP-even Higgs boson is NP-like.
In the opposite case, when $h^0$ is the SM-like (case a), the matching between the
full and effective theories yields
\begin{equation}
C_2 = 0, \hspace{0.5cm} \frac{C_1}{\Lambda_H^2} = \frac{1}{2}
\left(\frac{\lambda_2}{m_{h^0}^2} -
\frac{\lambda_3 \tan{\beta}}{m_{H^0}^2} \right)
\label{t6}
\end{equation}
In deriving (\ref{t6}), we used, for $\tan{\beta} \gg 1$ and $\xi \approx \pi/2 - \beta$,
\begin{equation}
\lambda_{h^0 \Phi} \approx \lambda_2, \hspace{0.5cm}
\lambda_{H^0 \Phi} \approx \lambda_3 \label{t17}
\end{equation}
It is not hard to see from our further analysis that omitted $O(1/\tan{\beta})$ terms in
Eqs.~(\ref{t17}) are not essential in this case.

The first term in the expression for $C_1/\Lambda_H^2$ is due to the SM-like Higgs
exchange. As expected, it has no enhancement factor - thus, as it was discussed
in Section~\ref{s8}, contribution of this term to
the $\Upsilon(3S) \to \Phi \Phi \gamma$
rate is by (at least) an order of magnitude lower than the present experimental
sensitivity. We may further disregard the Dark Matter interaction
with the lightest CP-even Higgs $h^0$.
Then, we may rewrite matching conditions (\ref{t6}) in a more
transparent form:
\begin{equation}
C_1 = \frac{- \lambda_3 \tan{\beta}}{2} \ , \hspace{0.5cm}
C_2 = 0 \ , \hspace{0.5cm} \Lambda_H = m_{H^0} \label{t22}
\end{equation}

In other words, we restrict ourselves by considering the contribution
to $\Upsilon(3S) \to \Phi \Phi \gamma$ amplitude due to exchange of the heaviest
(NP-like) Higgs boson only. This contribution is enhanced
by $\tan{\beta}$ factor, coming from
$\bar{b} H^0 b$ Yukawa interaction coupling in (\ref{t13}) (if $\xi \approx
\pi/2 - \beta$).
The remarkable
feature of the considered scenario is
that even though $\Upsilon(3S) \to \Phi \Phi \gamma$ transition
is generated by
exchange of the heaviest CP-even Higgs boson, the decay branching ratio for a
certain range of $H^0$ mass is well
within the reach of the
present experimental sensitivity, due to $\tan^2{\beta}$ enhancement.
As a consequence, one may derive
constraints on the coupling $\lambda_3$, as a function of $m_{H^0}$, $m_\Phi$
and $\tan{\beta}$.

Indeed, using Eqs.~(\ref{t22}), one may rewrite bound (\ref{g27}) on
$\sqrt{C_1^2 + C_2^2}$ as
\begin{equation}
|\lambda_3| <
\left(\frac{17.4}{\tan{\beta}} \right)
\left(\frac{m_{H^0}}{160 GeV} \right)^2
f^{-1/2}(x_\Phi) \label{t18}
\end{equation}
Choosing $m_{H^0} = 160~GeV$ as a reference value is not accidental.
Within the general type~II 2HDM, theoretical upper bound restricts the SM-like
Higgs to be less than 180 GeV \cite{26}. Also,
the SM Higgs mass interval $(160 - 170)~GeV$ has
been recently excluded with 95\% C. L. by the CDF and D0 data \cite{33}.
Thus, within type~II 2HDM, above 160 GeV, the CP-even Higgs boson is
presumably the heaviest one and NP-like.

\begin{figure}[t]
\includegraphics[width=0.49\textwidth]{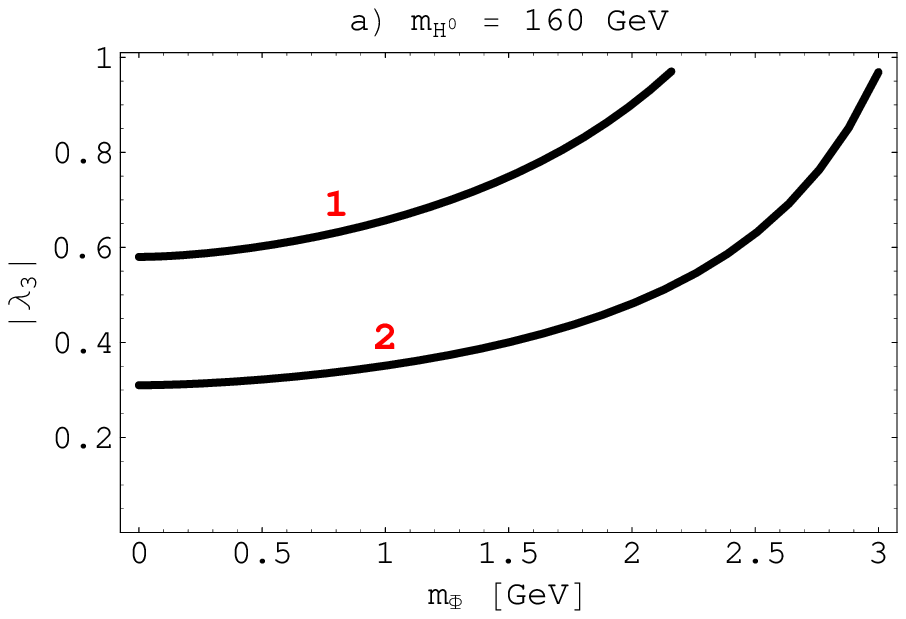} \hfill
\includegraphics[width=0.49\textwidth]{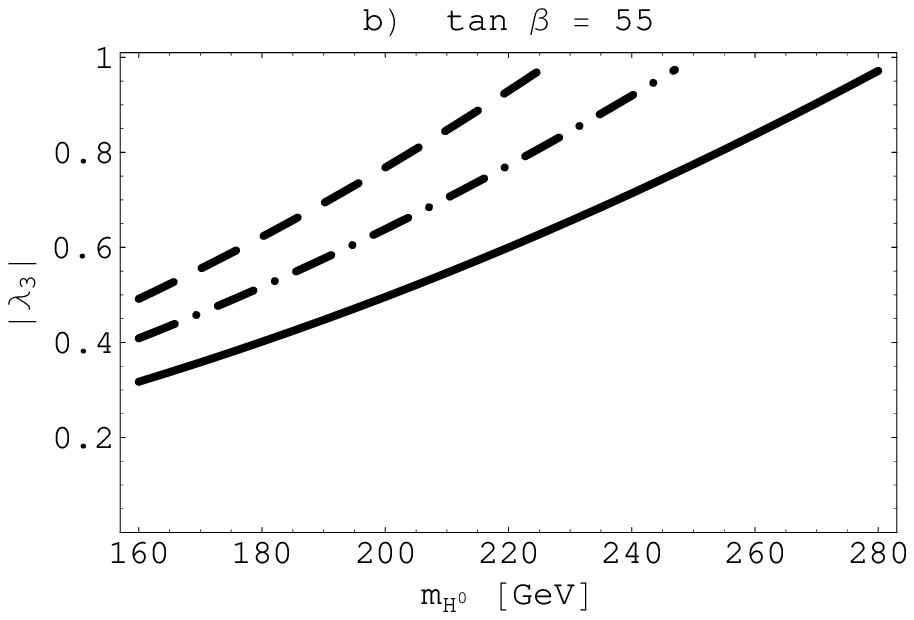}
\caption{Upper bound on $|\lambda_3|$, a) as a function of $m_\Phi$ for
$m_{H^0} = 160~GeV$ and $\tan{\beta} = 30$ (line 1), $\tan{\beta} = 55$
(line 2), b) as a function of $m_{H^0}$ for $\tan{\beta} = 55$ and
$m_\Phi = 100~MeV$ (solid line), $m_\Phi = 1.5~GeV$ (dashed-dotted line),
$m_\Phi = 2~GeV$ (dashed line).}
\label{f5}
\end{figure}

As one can see from Eq.~(\ref{t18}) and Fig.~\ref{f5}(a),
for $m_{H^0} = 160~GeV$ and $\tan{\beta} = 30$, $\lambda_3$ is constrained to
be of order of the SM weak coupling or smaller ($|\lambda_3| \lesssim 0.65$), if
$m_\Phi \lesssim 1~GeV$. Also, for the same choice of the Higgs mass and
$\tan{\beta}$, $|\lambda_3|$ is to be less than one, if the WIMP mass is less than
2 GeV.
Bound on $|\lambda_3|$ is significantly more rigorous for higher values of
$\tan{\beta}$. For instance, if choosing
$\tan{\beta} = m_t(m_t)/m_b(m_t) \approx 55$, one gets
$|\lambda_3| \lesssim 0.35$
and $|\lambda_3| < 0.5$ for $m_\Phi \lesssim 1~GeV$ and $m_\Phi \simeq 2~GeV$
respectively.

The restrictions on $|\lambda_3|$ are essential also for higher values of
the heaviest CP-even Higgs mass: for $\tan{\beta} = 55$, they are still
of interest
up to $m_{H^0} \simeq 280~GeV$, as one can see from Fig.~\ref{f5}(b). Our
analysis may be spread for the $m_{H^0} < 160~GeV$ range as well,
leading to more rigorous constraints than those in Fig.~\ref{f5}.

Recall that bound on $|\lambda_3|$ is derived for the first time.
It is also worth noting that constraints on $|\lambda_3|$ imply also
those on the heaviest CP-even Higgs invisible decay rate, if that Higgs is
NP-like. As it follows from Eq.~(\ref{t17}), $\lambda_3$ is the $H^0 \Phi \Phi$
interaction coupling in this case.
Yet, as in the case (b), bound (\ref{t18}) does not
seem to have an impact on the DM annihilation rate (which is $\tan^2{\beta}$
enhanced). Vice versa, DM annihilation processes (and possibly scattering
off nucleons) may lead to additional constraints on $\lambda_3$. Study of
these processes, however, goes beyond the scope of the present paper.

Thus, within type~II 2HDM with a scalar Dark Matter, for large $\tan{\beta}$ scenario,
$\Upsilon$ meson decay into a Dark Matter particles pair and a photon, $\Upsilon \to
\Phi \Phi \gamma$, may be used to derive essential constraints on the parameters
of the model, which otherwise cannot be tested by B meson decays with invisible
outcoming particles.

\section{Conclusions and Summary}
\label{s3}
Thus, spin-0 Dark Matter production in $\Upsilon$ meson decays has been investigated.
We restricted ourselves by consideration of the models where the decays occur due
to exchange of heavy non-resonant degrees of freedom. Both the scenarios with
a complex scalar DM field and those with DM particle being its own antiparticle
have been analyzed.

We performed our calculations within low-energy effective theory, integrating out
heavy degrees of freedom. This way we derived model-independent formulae for
the considered branching ratios. We used these formulae to confront our
theoretical predictions with existing experimental data on invisible
$\Upsilon$ decays, both in a model-independent way and within particular
models. It has been shown that within the considered class of models, DM
production rate in $\Upsilon$ decays is within the reach of the present
experimental sensitivity. Thus, $\Upsilon$ meson decays into Dark Matter,
with or without a photon emission, may be used to constrain the models
with  a GeV or lighter spin-0 DM. In particular, within the mirror fermion models,
using the existing BaBaR constraint on the $\Upsilon(1S) \to invisible$
mode, we derived for the first time bounds on the parameters of the model
that otherwise could not be tested by other DM search processes.

Experimental constraints on the other mode, $\Upsilon(3S) \to \gamma + invisible$, are
derived assuming that Dark Matter is produced by exchange of a light
resonant scalar state. Within the scenarios with non-resonant DM
production, these constraints may be used only to make preliminary
estimates of possible bounds on the parameters of the models. Yet,
those estimates show that these bounds may be rigorous enough; besides,
they are derived within the least restrained presently light scalar DM
scenarios.
Our goal is thus to encourage the experimental groups to analyze
the experimental data on $\Upsilon \to \gamma + invisible$ also
for the case of non-monochromatic photon emission and spin-0
invisible states.

So, from our analysis one may conclude that Dark Matter production in
$\Upsilon$ meson decays may serve as an interesting alternative to
commonly used DM search methods, capable of providing a valuable
information on DM particles, if those turn to have a mass of
the order of a few GeV or smaller.

\acknowledgments

The author is grateful to Yu. Kolomensky, A. Blechman,  A. A. Petrov, B. McElrath and
J. Cao for valuable
comments and stimulating discussions. The author thanks A. Badin for verifying some of
derivations. \\

This work has been supported by NSF Grant No.~PHY-0547794 and DOE Grant No.~DE-FGO2-96ER41005.



\end{document}